\documentclass[apjl,a4paper,12pt,useAMS]{emulateapj}
\usepackage{amssymb}
\usepackage{graphicx}

\begin{document}
\title{A statistical study of super-luminous supernovae in the magnetar engine model and implications for their connection with gamma-ray bursts and hypernovae}
\author{Yun-Wei~Yu$^{1,2}$, Jin-Ping Zhu$^{1}$, Shao-Ze Li$^{1}$, Hou-Jun L\"u$^{3,4}$, Yuan-Chuan Zou$^{5}$}

\altaffiltext{1}{Institute of Astrophysics, Central China Normal
University, Wuhan 430079, China, {yuyw@mail.ccnu.edu.cn}}
\altaffiltext{2}{Key Laboratory of Quark and Lepton Physics (Central
China Normal University), Ministry of Education, Wuhan 430079,
China}\altaffiltext{3}{GXU-NAOC Center for Astrophysics and Space
Sciences, Department of Physics, Guangxi University, Nanning 530004,
China} \altaffiltext{4}{Guangxi Key Laboratory for Relativistic
Astrophysics, Nanning, Guangxi 530004, China}\altaffiltext{5}{School of Physics, Huazhong University of Science and Technology, Wuhan, China}

\begin{abstract}
By fitting the bolometric light curves of 31 super-luminous
supernovae (SLSNe) with the magnetar engine model, we derive the
ejecta masses and magnetar parameters for these SLSNe. The lower
boundary of magnetic field strengths of SLSN magnetars can be set
just around the critical field strength $B_{\rm c}$ of electron
Landau quantization. In more details, SLSN magnetars can further be
divided into two subclasses of magnetic fields of $\sim(1-5)B_{\rm
c}$ and $\sim(5-10)B_{\rm c}$, respectively. It is revealed that
these two subclasses of magnetars are just associated with the
slow-evolving and fast-evolving bolometric light curves of SLSNe. In
comparison, the magnetars harbored in gamma-ray bursts (GRBs) and
associated hypernovae are usually inferred to have much higher
magnetic fields with a lower boundary about $\sim10B_{\rm c}$. This
robustly suggests that it is the magnetic fields that play the
crucial role in distinguishing SLSNe from GRBs/hypernovae. The
rotational energy of SLSN magnetars are found to be correlated with
the masses of supernova ejecta, which provides a clue to explore the
nature of their progenitors. Moreover, the distribution of ejecta
masses of SLSNe is basically intermediate between those of normal
core-collapse supernovae and hypernovae. This could indicate an
intrinsic connection among these different stellar explosions.
\end{abstract} 

\section{Introduction}
Superluminous supernovae (SLSNe) are an unusual type of supernovae
intrinsically bright with a peak absolute magnitude of $M<-21$,
which are about  $10-100$ times brighter than normal supernova
events (Gal-Yam et al. 2012). The event rate of SLSNe was estimated
to several tens to a hundred times per year per cubic gigaparsecs at
redshift $z\sim 1$, which increases with redshift in a manner
consistent with that of the cosmic star formation history (Prajs et
al. 2017). As usual, SLSNe can be divided observationally into two
classes according to the detection of hydrogen in their spectra,
i.e., hydrogen-poor SLSNe I (Quimby et al. 2011) and hydrogen-rich
SLSNe II (Ofek et al. 2007; Smith et al. 2007). SLSNe I commonly
have a blue continua at peak and a distinctive W-shaped feature
identified as O II  at early epochs. A few tens of days after peak,
their spectral evolution becomes very similar to normal or
broad-lined Type Ic supernovae  (Pastorello  et  al. 2010). SLSNe II
have a blue continua too at maximum light and, more importantly,
some clear Balmer features exist in their spectra. The current
observed number of SLSNe I is apparently larger than that of SLSNe
II. In addition, for some particular SLSNe (e.g. iPTF13ehe), their
spectral type could even evolve from one to another as the fading of
the emission (Yan et al. 2015). With one exception (PTF 10uhf), the
host galaxies of SLSNe I are always found to be low-mass and
low-metallicity ($\lesssim 0.5Z_{\odot}$), whereas those of SLSNe II
are usually all over the entire range of galaxy masses and
metallicities (Neill et al. 2011; Lunnan et al. 2014; Leloudas et
al. 2015; Chen et al. 2016a; Angus et al. 2016; Perley et al. 2016).

The energy sources powering SLSNe are puzzling, while the
traditional scenario of radioactive decays of heavy elements is
seriously challenged by the unusually high luminosity. The total
radiated energy of a typical SLSN is about a few times $10^{51}$
erg. If this radiation is mainly powered as usual by the radioactive
chain $\rm ^{56}Ni\rightarrow^{56}Co\rightarrow^{56}Fe$, then an
extremely large amount (several to several tens of solar masses) of
radioactive nickel is required to be produced during the supernova
explosion, which is unfortunately impossible for normal supernova
nucleosynthesis (e.g. Umeda \& Nomoto 2008). Nevertheless, such a
high mass of nickel could still be produced during some unusual
supernova explosions that are triggered due to electron-positron
pair-production instability (Barkat et al. 1967; Heger \& Woosley
2002). Gal-Yam et al. (2009) firstly used this pair-instability
supernova (PISN) model to explain the light curve of hydrogen-poor
SN 2007bi, which decays very slowly at late phase being consistent
with the decay of radioactive nickel and cobalt (cf. see Kozyreva \&
Blinnikov 2015). In contrast, a counterview was argued by Nicholl et
al. (2013) as that the PISN model could be disfavored by the fast
rising of two SN 2007bi-like SLSNe. In any case, it is expected
that PISNe tend to appear at relatively high redshifts (McCrum et
al. 2014), which can significantly suppress their detection
probability at relatively near distances. Therefore, even if a few
SLSNe can indeed be ascribed to PISNe, most other SLSNe (in
particular, the fast evolving ones) inevitably need some other
powerful energy sources alternative to radioactivities.

On one hand, the broad-lined features in SLSN spectra indicate that
the supernova ejecta moves at a very high speed and thus carries
huge kinetic energy. Therefore, it is natural to consider that the
ejecta can be heated by consuming the kinetic energy through shock
interaction with circum-stellar material (CSM; Smith \& McCray 2007;
Moriya et al. 2011, 2013; Chevalier \& Irwin 2011; Ginzburg \&
Balberg 2012), if the CSM is dense and extended sufficiently. It is
interesting to mention that such opaque CSM could sometimes be
produced by violent pulsation of a massive star triggered by pair
instability, before the final disruption of the star (i.e., the
pulsational PISN model; Woosley et al. 2007; Chatzopoulos \& Wheeler
2012). Observationally, the CSM-interaction model can be strongly
supported by the existence of intermediate and narrow Balmer
emission lines in the spectra of some SLSNe II termed SLSN IIn.
These lines are formed due to recombination of ionized CSM.

On the other hand, usually for hydrogen-poor SLSNe I, it is
considered that their emission could be associated with a long-lived
central engine, which can release energy persistently. In this case,
the supernova ejecta can be heated gradually by absorbing the
engine-released energy. Very recently, Inserra et al. (2016a)
discovered an axi-symmetric ellipsoidal configuration for the ejecta
of SN 2015bn. Such a geometry, which is similar to those of
hypernovae associated with gamma-ray bursts (GRBs), strongly
indicates the significant influence of a central engine. To be
specific, the energy release from a central engine could be due to
spin-down of a millisecond magnetar (Ostriker \& Gunn 1971; Woosley
et al. 2010; Kasen et al. 2010, 2016; Moriya et al. 2016; Chen et
al. 2016b) or/and due to feedback of fallback accretion (Dexter \&
Kasen 2013). In comparison, the former scenario was much more
adopted in literature, probably because the allowable ranges of
luminosities and timescales of magnetar spin-down are much wider
than those of fallback accretion (e.g. Yu \& Li 2016).

In view of its obvious advantages, the magnetar engine model has
been widely employed to explain SLSN emission either individually
(e.g., Dessart et al. 2012; Nicholl et al. 2013; Howell et al. 2013;
McCrum et al. 2014; Dai et al. 2016) or in bulk (e.g., Inserra et
al. 2013, 2016b; Chatzopoulos et al. 2013; Wang et al. 2015; Nicholl
et al. 2015a). These works can usually go to success and account for
observations usually better than the models of CSM interaction and
of radioactive decays. In this paper we try to carry out such a work
with the maximum sample possible, in order to reveal some
statistical properties of magnetar-powered SLSNe and the
corresponding magnetars, and compare the results with statistics of
long GRBs and their associated hypernovae. The paper is organized as
follows. In Section 2 the magnetar engine model is introduced
briefly. In Section 3 we collect observational data of SLSNe from
literature and fit the bolometric light curves of the SLSNe with the
magnetar model. The statistics of the obtained model parameters are
given in Section 4. In Section 5, we discuss the possible
connections and differences between SLSNe and GRBs/hypernovae as
well as normal Type Ic broad-lined (Ic-BL) supernovae. Summary and
conclusion are given in Section 6.

\section{Magnetar engine model}\label{SectModel}
Similar to previous statistical works of SLSNe (e.g., Nicholl et al.
2015a), a simple semi-analytical model is adopted in this paper to
calculate the emission of magnetar-powered SLSNe. Following Kasen \&
Bildsten (2010), the bolometric luminosity of a supernova can be
roughly determined by the following formula:
\begin{eqnarray}
L_{\rm sn}={cE_{\rm int}\over R_{\rm }\tau}\left(1-e^{-\tau}\right),
\end{eqnarray}
which is derived according to the heat diffusion in the supernova
ejecta, where $c$ is the speed of light, $E_{\rm int}$ is the total
internal energy of the ejecta, $R$ is the ejecta radius, and $\tau$
is the optical depth. For an ejecta mass of $M_{\rm ej}$, we can
estimate $\tau=3\kappa M_{\rm ej}/4\pi R^2$, where $\kappa$ is the
opacity. For $\tau\gg 1$ the above equation reads $L_{\rm sn}=
cE_{\rm int}/( R\tau)$, while $L_{\rm sn}= cE_{\rm int}/R$ for $\tau
\ll1$ (e.g., see Kotera et al. 2013). For simplicity, a constant
value of $\kappa=0.1\rm cm^{2}g^{-1}$ is adopted in following
calculations, which is typical and reasonable for SLSNe (Inserra et
al. 2013; Nicholl et al. 2015a). By considering of the energy
conservation of the ejecta, the evolution of its internal energy can
be calculated by
\begin{eqnarray}
\frac{ d E_{\rm int}}{d t} =  L_{\rm sd} - L_{\rm sn}- 4\pi R_{\rm
}^2 pv_{\rm }  ,\label{Eint}
\end{eqnarray}
where $t$ is the time, $L_{\rm sd}$ is the energy injection rate due
to spin-down of a magnetar, $p$ is the pressure that can be related to the internal
energy by $p={1\over3}(E_{\rm int}/{4\over3}\pi R_{\rm }^3)$, and
$v_{\rm }={dR_{\rm }/ dt}$ is the expansion speed of the ejecta. The
term $4\pi R_{\rm }^2 pv_{\rm }dt$ represents the adiabatic cooling
of the ejecta, which energy is used to accelerate the ejecta.
Therefore, the dynamical equation can be written as
\begin{eqnarray}
{dv_{\rm }\over dt}={4\pi R_{\rm }^2p\over M_{\rm ej}}.
\end{eqnarray}
For a dipolar magnetic field of polar strength of $B_{\rm p}$, the
spin-down luminosity of a magnetar as a function of time can be
expressed by the magnetic dipole radiation formula:
\begin{eqnarray}
L_{\rm sd}=L_{\rm sd,i}\left(1+{t\over t_{\rm
sd}}\right)^{-2}\label{Lp}
\end{eqnarray}
with an initial value of $L_{\rm sd,i}=10^{47}~B_{\rm
p,14}^{2}P_{\rm i,-3}^{-4}\rm ~erg~s^{-1}$ and a spin-down timescale
of $t_{\rm sd}=2\times10^{5}~B_{\rm p,14}^{-2}P_{\rm i,-3}^{2}\rm
s$, where $P_{\rm i}$ is the initial spin period of the magnetar.
Hereafter the conventional notation $Q_x=Q/10^x$ is adopted in cgs
units.

The above equations can be solved numerically by taking an initial
kinetic energy for the ejecta and assigning values for the model
parameters: $L_{\rm sd,i}$, $t_{\rm sd}$, and $M_{\rm ej}$. As a
result, the bolometric luminosity of a supernova can be obtained as
a function of time\footnote{The solution is usually approximated by
the following integral (Arnett 1982):
\begin{eqnarray}
L_{\rm sn}(t)=e^{-\left({t\over t_{\rm d}}\right)^2}\int_0^t 2L_{\rm
sd}(t'){t'\over t_{\rm d}}e^{\left({t'\over t_{\rm
d}}\right)^2}{dt'\over t_{\rm d}},
\end{eqnarray}
where the dynamical evolution of the supernova ejecta is ignored and
its heat diffusion timescale is given by $t_{\rm d}=(3\kappa M_{\rm
ej}/4\pi v_{\rm f}c)$. The final speed of ejecta $v_{\rm f}$ can be
determined from ${1\over2}M_{\rm ej}v_{\rm f}^2=\left(E_{\rm
kin,i}+L_{\rm sd,i}t_{\rm sd}-E_{\rm rad}\right)$, where $E_{\rm
rad}$ is the energy release via radiation.}. Statistics of normal
core-collapse supernovae showed that their kinetic energies are
concentrated within the range from $10^{51}$ erg to several times
$10^{51}$ erg (Lyman et al. 2014). It is considered here that the
initial trigger processes of SLSN explosions could be similar to
those of normal core-collapse supernovae. In this case, the SLSN
ejecta could obtain a similar initial kinetic energy on the order of
$10^{51}$ erg, which is taken in our calculations. It should be
emphasized that the kinetic energies of most SLSN ejecta can quickly
become much larger than this initial value because of the energy
injection from magnetar. Therefore, to a certain extent,
fittings to observational light curves would not be very
sensitive to the choice of the initial value of kinetic energy.

\section{Data collection and fittings}
Up to December 2016, dozens of SLSNe have been discovered by
different authors with different telescopes. The consistency of
these SLSNe with the magnetar engine model had been individually
inspected in literature by modeling their light curves as well as
their temperature and velocity evolutions. On one hand, the magnetar
model is most favored by SLSNe I. Therefore, we have primarily
collected 27 SLSNe I whose bolometric light curves had been given in
the literature. Here, two unique transients (i.e. PS1-10afx and
ASASSN-15lh), which were classified to SLSNe I initially (Chornock
et al. 2013; Dong et al. 2016), are not included in our sample,
because the former is likely to be a lensed Type Ia supernova
(Quimby et al. 2014) and the latter has very unusual behaviors
(Godoy-Rivera et al. 2017) and is even suggested to be a tidal
disruption event from a Kerr black hole (Leloudas et al. 2016). On
the other hand, for SLSNe II, part of them (i.e., SLSNe IIn such as
CSS100217, SN 2006tf, SN 2006gy, and SN 2008am) can be undoubtedly
ascribed to an ejecta-CSM interaction. However, as pointed out by
Inserra et al. (2016b), some other SLSNe II, which have only broad
H$\alpha$ features in their early spectra, could still well be
explained by the magnetar engine model. Therefore, such four
broad-lined SLSNe II (i.e., CSS121015, PS15br, SN 2008es, SN 2013hx)
will also be taken into account in our statistics, in order to
maximize the sample number.

\begin{figure}
\centering\resizebox{0.3\hsize}{!}{\includegraphics{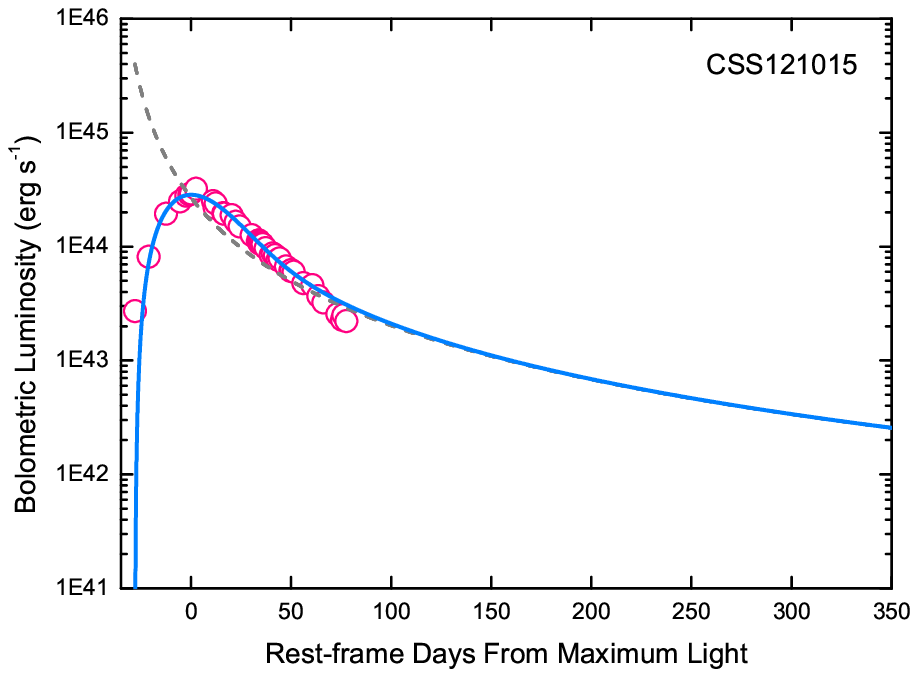}}\resizebox{0.3\hsize}{!}{\includegraphics{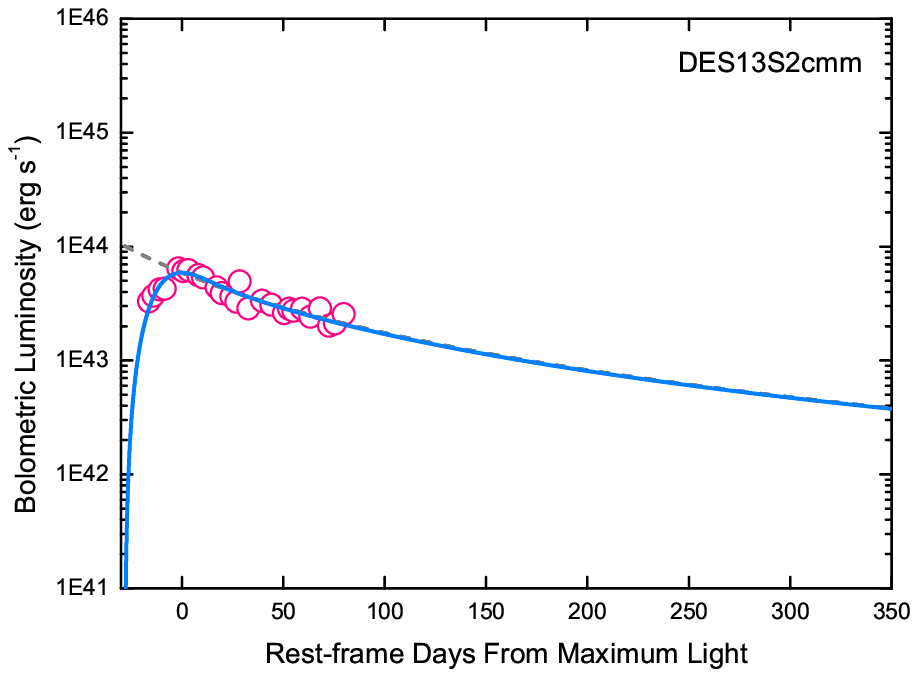}}\resizebox{0.3\hsize}{!}{\includegraphics{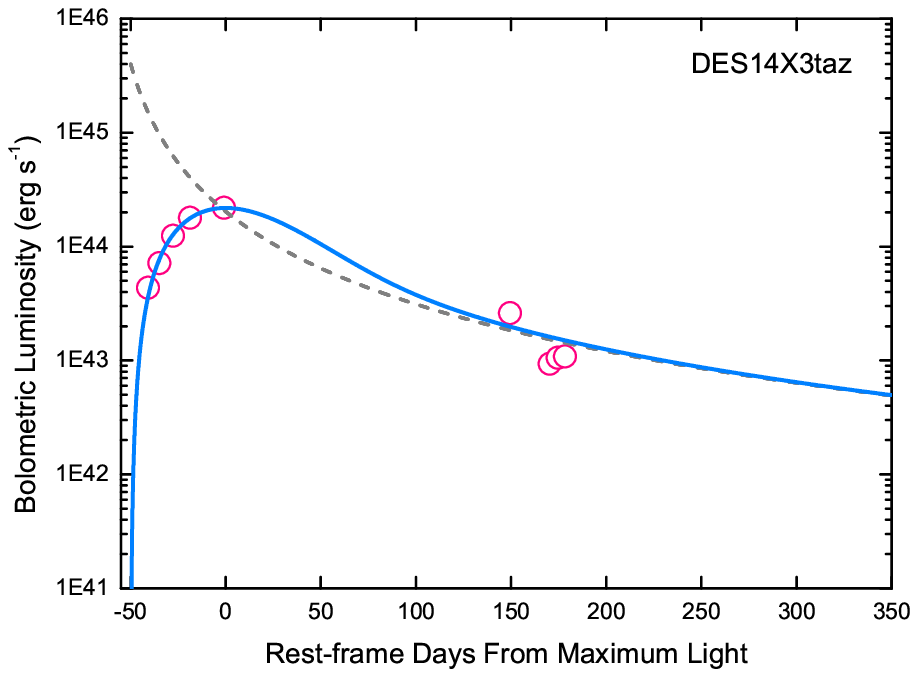}}
\centering\resizebox{0.3\hsize}{!}{\includegraphics{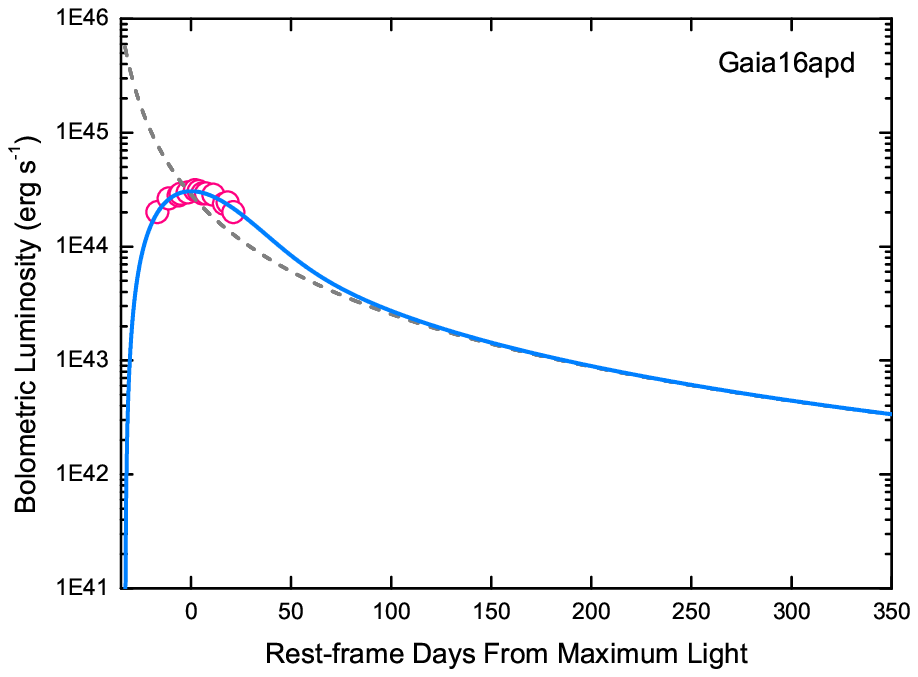}}\resizebox{0.3\hsize}{!}{\includegraphics{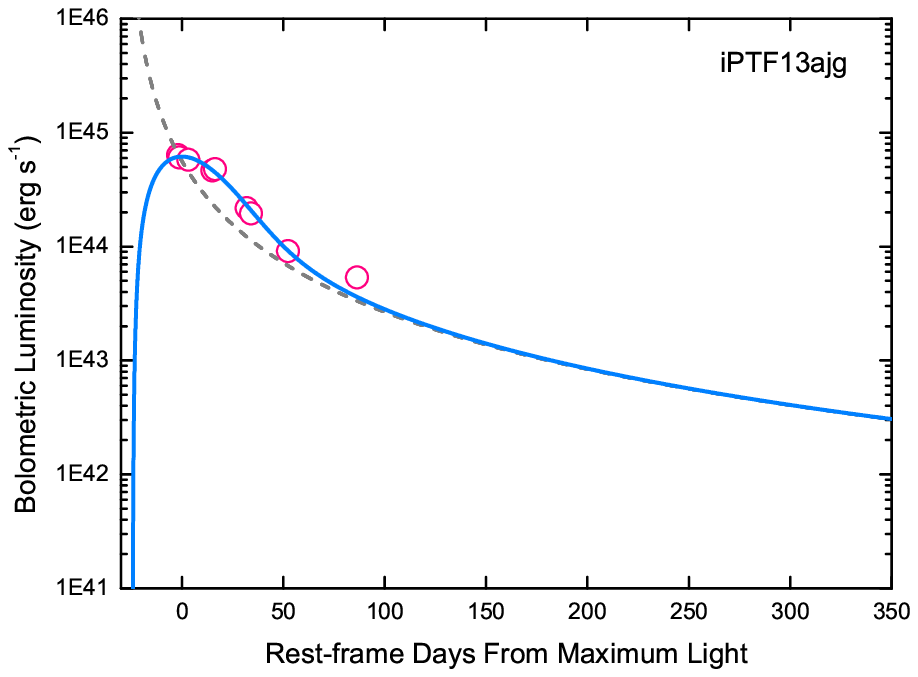}}\resizebox{0.3\hsize}{!}{\includegraphics{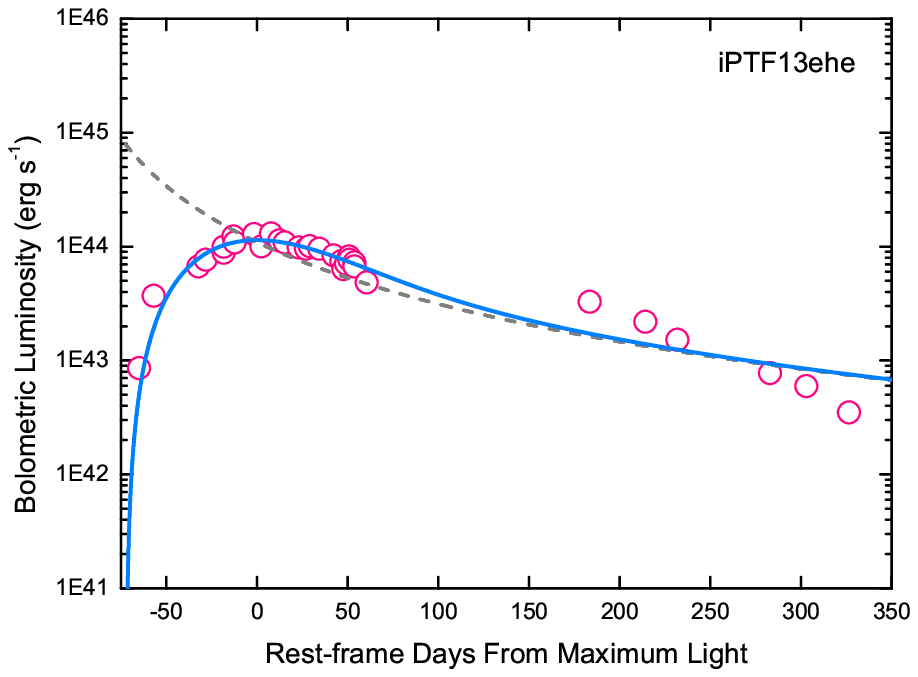}}
\centering\resizebox{0.3\hsize}{!}{\includegraphics{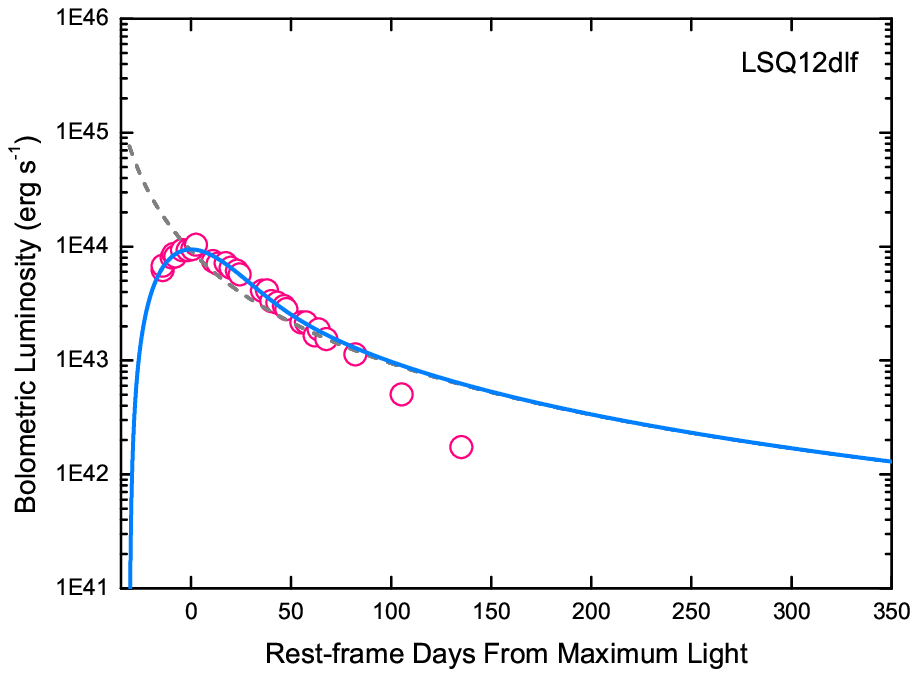}}\resizebox{0.3\hsize}{!}{\includegraphics{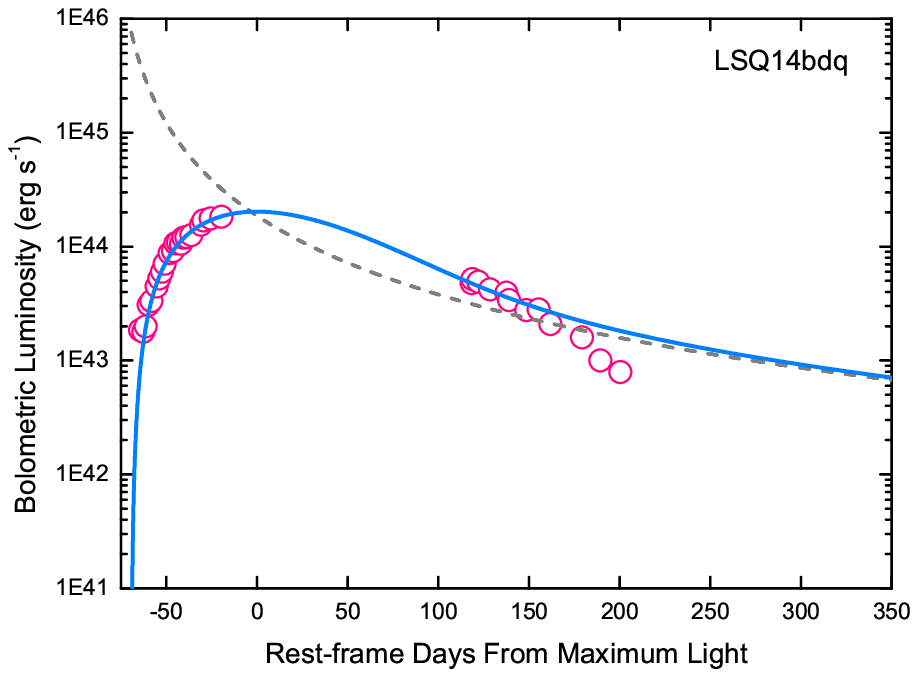}}\resizebox{0.3\hsize}{!}{\includegraphics{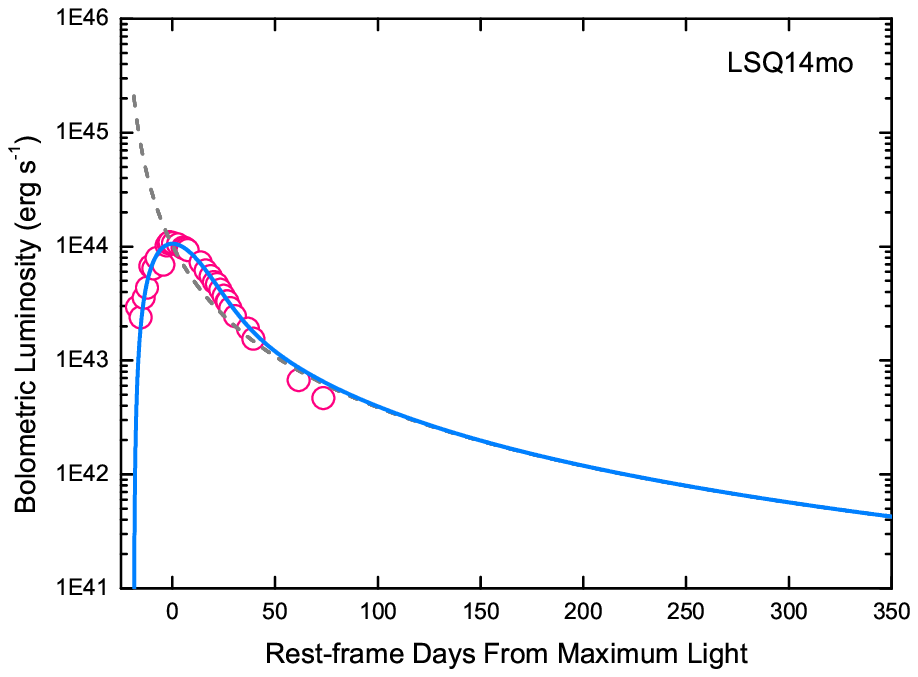}}
\centering\resizebox{0.3\hsize}{!}{\includegraphics{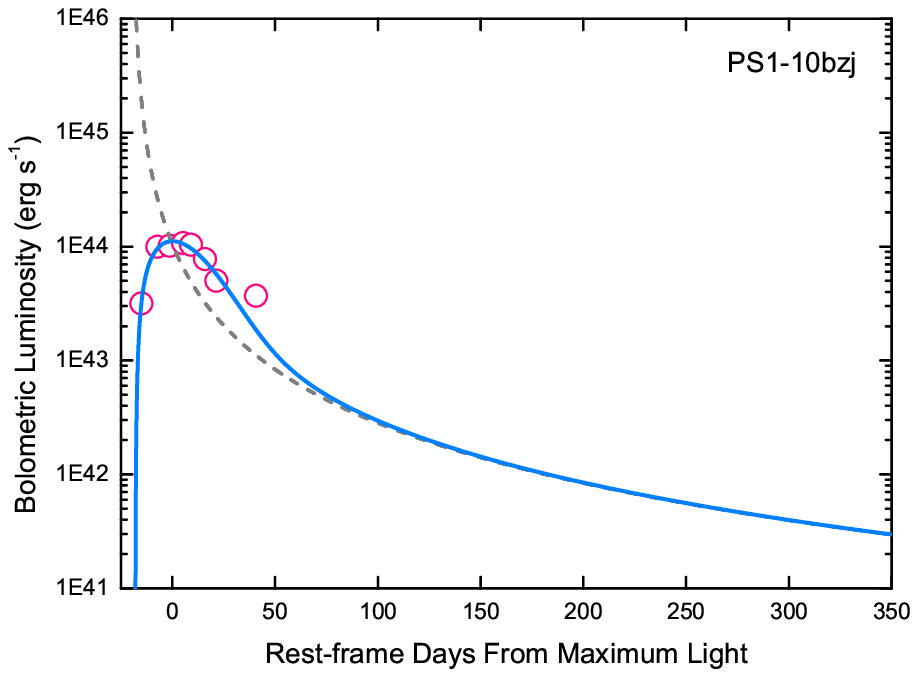}}\resizebox{0.3\hsize}{!}{\includegraphics{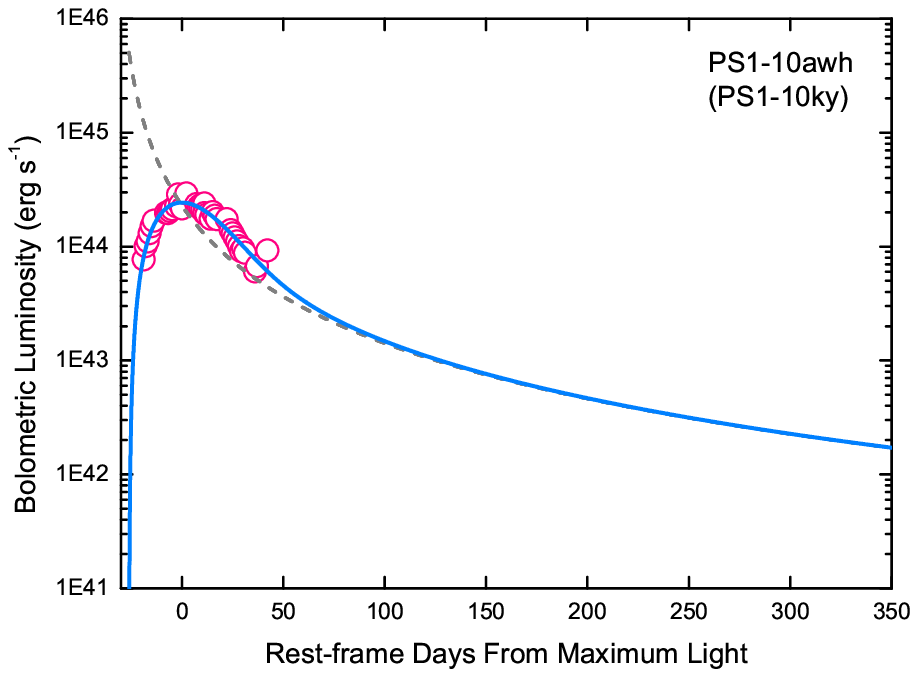}}\resizebox{0.3\hsize}{!}{\includegraphics{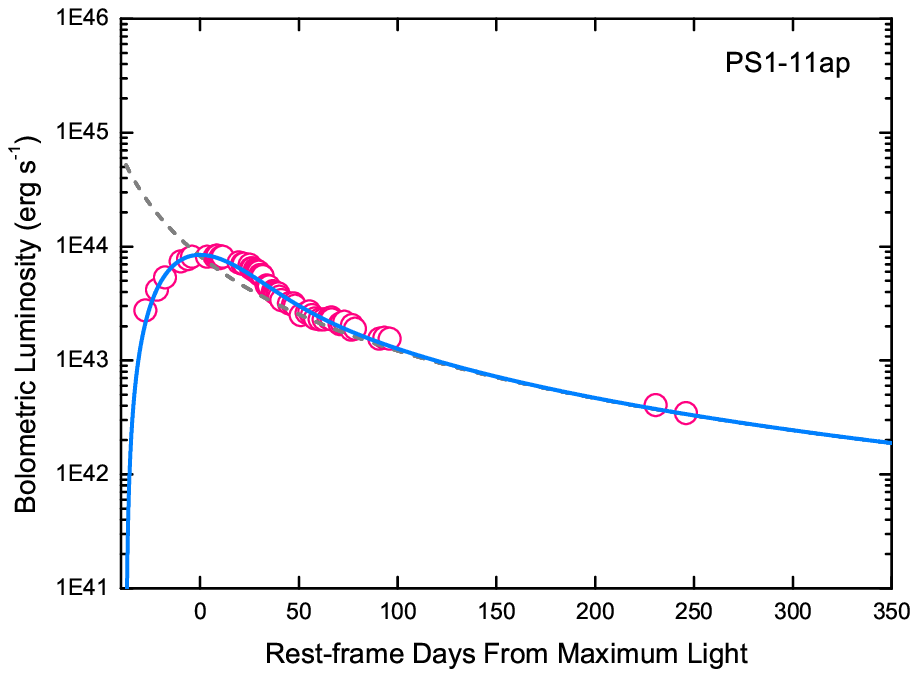}}
\centering\resizebox{0.3\hsize}{!}{\includegraphics{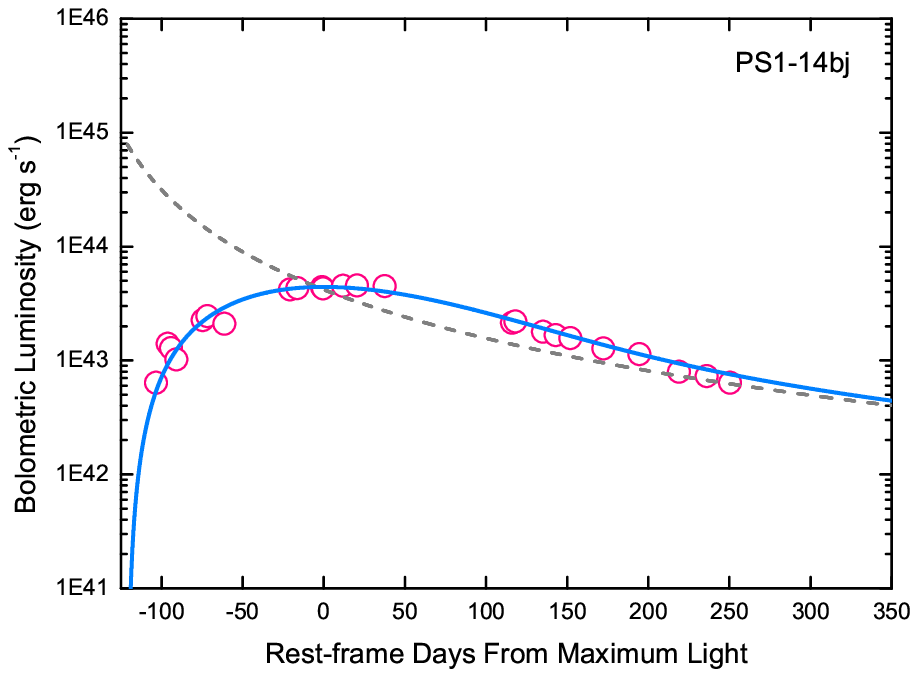}}\resizebox{0.3\hsize}{!}{\includegraphics{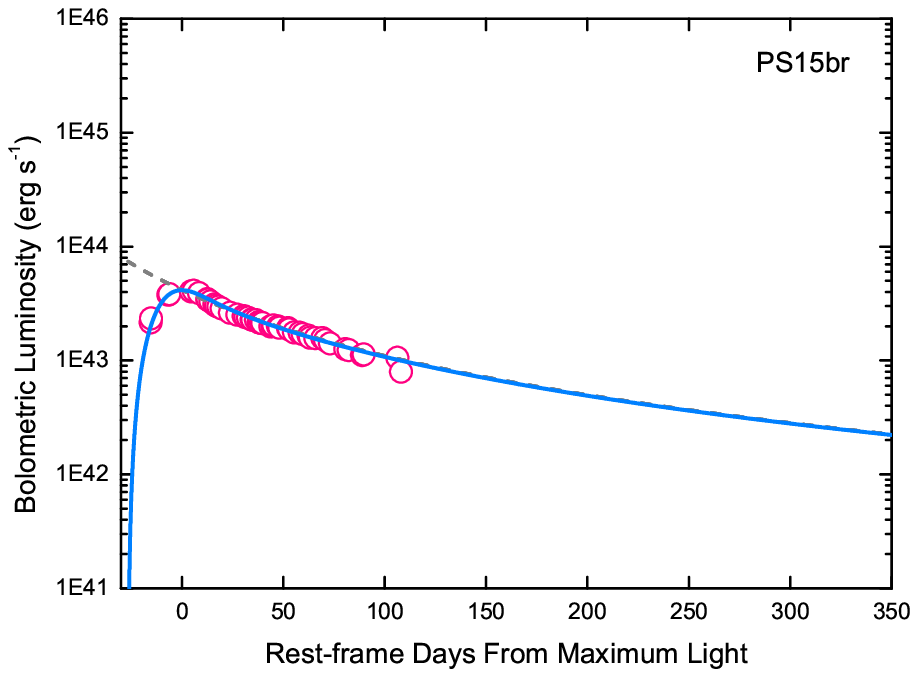}}\resizebox{0.3\hsize}{!}{\includegraphics{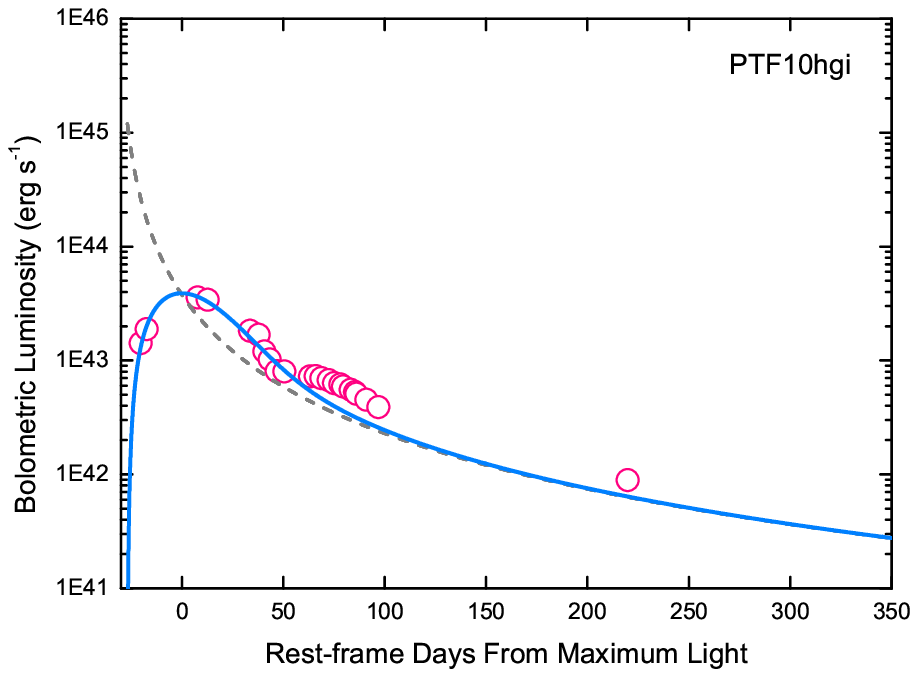}}
\caption{Fittings to SLSN light curves in the magnetar engine model
(solid line), where individual characteristics in late observations
are not taken into account. The dashed line represents the spin-down
luminosity of the magnetar.}\label{figfittings}
\end{figure}

The basic information of the 31 selected SLSNe is listed in Table
\ref{tabobspara} including supernova names, spectral types,
coordinates, redshifts, and references. From the listed references,
we take the light curves of these SLSNe and present them in Figure
\ref{figfittings}. These light curves are regarded as bolometric as
reported in the references and no further correction is made for
them in this paper. Technically, for a statistical study, the
parameter values of these SLSNe in the magnetar engine model can
also be taken directly from the references. Nevertheless, here we
choose to refit the bolometric light curves of all 31 SLSNe in a
united method as described in Section \ref{SectModel}, just for a
general self-consistency. The obtained values of model parameters
(i.e., $L_{\rm sd,i}$, $t_{\rm sd}$, and $M_{\rm ej}$) are listed in
Table \ref{tabfitpara}, which are generally consistent with the
results presented in previous literature (e.g. Chatzopoulos et al.
2013). The fitting results are presented in Figure
\ref{figfittings}. As shown, the light curves can well be fitted by
the magnetar engine model during early times, including the rising
and decreasing phases generally until $\sim100-200$ days after peak.

In despite of the good modelings of the light curve peaks, at late
time the fitting curves could sometimes deviate from the
observational data. Firstly, for some SLSNe (e.g., LSQ 12dlf, LSQ
14bdq, and PTF 12dam), the theoretical light curves could become
higher and higher than the late data, which is probably due to our
neglecting of the suppression of energy absorption
efficiency\footnote{The energy injected into supernova ejecta from a
magnetar is usually considered to be in the form of high-energy
photons. As the expansion of the ejecta at late times, its optical
depth for high-energy photons could decrease quickly, which leads a
remarkable fraction of injected high-energy photons to leak from the
ejecta (see Wang et al. 2015 and references therein). Therefore, the
energy absorption efficiency is suppressed. Meanwhile, this leaked
high-energy emission could provide a smoking-gun observational
signature for the magnetar-engine model (e.g. Kotera et al. 2013),
which is worth to be explored in current and future observations.}.
Secondly, as Wang et al. (2016) suggested, while the peak emission
is dominated by the engine contribution, the late emission of some
SLSNe could be partially contributed by shock interactions and
radioactivities. Finally, of more interests, some frequent
undulations could appear in the light curves of some SLSNe (e.g.,
PS1-11ap, PTF10hgi, and SN2015bn), while the underlying profile of
these light curves can be explained successfully with a continuous
energy injection from a magnetar. These undulations could be caused
by collisions of supernova ejecta with some clumping CSM (Nicholl et
al. 2016) or by some late intermittent flare activities of the
magnetar engine (Yu \& Li 2016). In any case, the primary purpose of
this paper is just to get a universal set of model parameters for a
large enough SLSN sample, which can basically be determined by
fitting the light curves around peak. Therefore, detailed emission
characteristics of specific SLSNe such as those mentioned above are
ignored in our fittings.

\section{Statistics of model parameters}
 \subsection{Magnetar parameters}\label{SectMag}
\begin{figure}
\centering\resizebox{0.5\hsize}{!}{\includegraphics{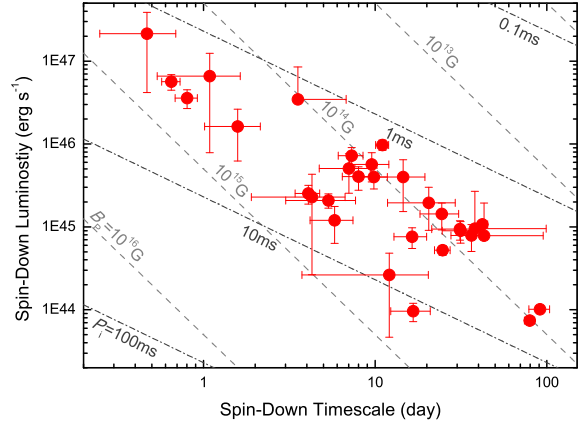}}
\caption{The spin-down luminosity of SLSN magnetars against the
spin-down timescale. The dashed and dash-dotted lines correspond to
different magnetic field strengths and initial spin periods as
labeled. } \label{figLsdtsd}
\end{figure}

In Figure \ref{figLsdtsd}, the spin-down timescales of the SLSN
magnetars are displayed against the initial values of spin-down
luminosity, where an anti-correlation seems to appear between these
two quantities. In fact, such an anti-correlation had also been
previously claimed for GRB magnetars, which exhibits as a
relationship between the luminosity and duration of the X-ray
afterglow plateaus of GRBs (Dainotti et al. 2008; Rowlinson et al.
2014). It is not difficult to understand this
relationship by combining the expressions of $L_{\rm sd,i}$ and
$t_{\rm sd}$. Specifically, for a fixed magnetic field we have
$L_{\rm sd,i,47}=5.0B_{\rm p,14}^{-2}t_{\rm sd,day}^{-2}$ or for a
fixed initial spin period $L_{\rm sd,i,47}=2.3P_{\rm
i,-3}^{-2}t_{\rm sd,day}^{-1}$. These expressions are presented in
Figure \ref{figLsdtsd} by the iso-$B_{\rm p}$ and iso-$P_{\rm i}$
lines. Therefore, for any arbitrary distributions of $B_{\rm p}$ and
$P_{\rm i}$, a seeming $L_{\rm sd,i}-t_{\rm sd}$ anti-correlation
can always be obtained, which is not surprising.

\begin{figure}
\centering\resizebox{0.7\hsize}{!}{\includegraphics{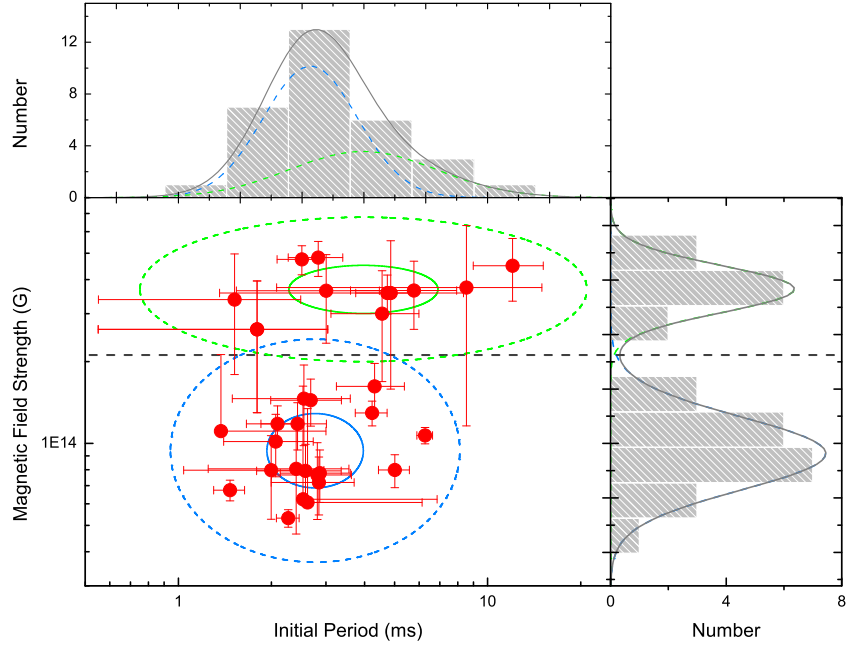}}
\caption{The magnetic filed strengths of SLSN magnetars against the
initial spin periods. The ellipses represent the 1$\sigma$ and
3$\sigma$ regions of the two subclasses defined by the
log-Gaussians. The upper and right panels show that the
distributions of magnetic field strengths and initial spin periods
can both be fitted by the sum (solid lines) of two log-Gaussians
(dashed lines). The parameters of the Gaussians are $\mu_1=13.96$,
$\sigma_1=0.13$ and $\mu_2=14.57$, $\sigma_2=0.09$ for the
distribution of $\log (B_{\rm p}/\rm G)$ and $\mu_1=0.43$ (2.7 ms),
$\sigma_1=0.15$ and $\mu_2=0.60$ (4.0 ms), $\sigma_2=0.24$ for the
distribution of $\log (P_{\rm i}/\rm ms)$.}\label{figBPi}
\end{figure}

It is undoubtedly of more scientific significance to investigate the
magnetic fields and spin periods of the SLSN magnetars, which can
provide more intrinsic information. The magnetic field strength and
initial spin period of a magnetar can be derived by
\begin{eqnarray}
B_{\rm p}=2\times10^{14}L_{\rm sd,i,47}^{-1/2}t_{\rm sd,5}^{-1}\rm G,
\end{eqnarray}
and
\begin{eqnarray}
P_{\rm i}=1.4\times10^{-3}L_{\rm sd,i,47}^{-1/2}t_{\rm sd,5}^{-1/2}\rm s.
\end{eqnarray}
Then, the obtained values of $B_{\rm p}$ and $P_{\rm i}$ of the SLSN
magnetars are listed in Table \ref{tabfitpara} and exhibited in
Figure \ref{figBPi}. By using the MCLUST software implementation by
Fraley \& Raftery
(2002)\footnote{http://www.stat.washington.edu/raftery/Research/mbc.html},
we fit the two-dimension distribution of the data in Figure
\ref{figBPi} by the Gaussian mixture model with the number of
components evaluated using the Bayesian information criterion. As a
result, it is found that the data distribution can well be fitted by
a mixture of two log-normal functions, as shown by the $1-\sigma$
and $3-\sigma$ contours in Figure \ref{figBPi}. The centering
parameter values of the two separated regions are $\langle B_{\rm
p}\rangle_{\rm SLSN1}=9.2\times10^{13}$ G, $\langle P_{\rm
i}\rangle_{\rm SLSN1}=2.7$ ms and $\langle B_{\rm p}\rangle_{\rm
SLSN2}=3.7\times10^{14}$ G, $\langle P_{\rm i}\rangle_{\rm
SLSN2}=4.0$ ms. To be specific, fittings to the histograms of $\log
B_{\rm p}$ and $\log P_{\rm i}$ are respectively presented in the
right and upper panels of Figure \ref{figBPi}. It is indicated that
the SLSN magnetars can empirically be divided into two subclasses by
a separating field strength of $2.2\times10^{14}$ G. In addition, we
would like to mention that, if we arbitrarily model some SLSNe that
are not included in our sample (e.g., CSS100217 and SN 2008am) by
the magnetar model, then the inferred parameters would be found to
be very far away from the 3-$\sigma$ edge of the above two
log-Gaussians. This indicates that these excluded SLSNe are indeed
not powered by a magnetar.

\begin{figure}
\centering\resizebox{0.5\hsize}{!}{\includegraphics{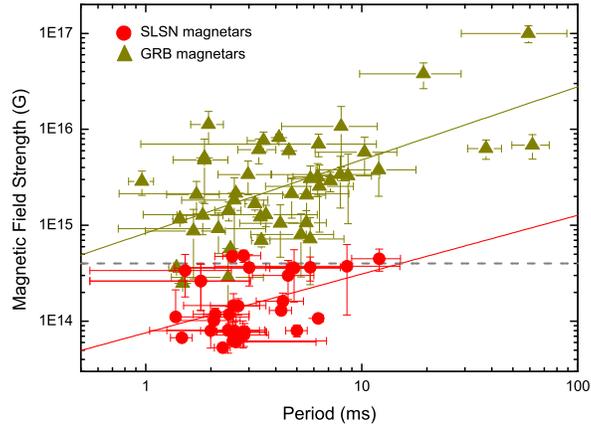}}
\caption{A comparison between the magnetar parameters of SLSNe and
long GRBs. The two data samples can be separated by the dashed line.
The solid lines represent a possible correlation between $B_{\rm p}$
and $P_{\rm i}$ of the two types of magnetars.} \label{figGRB}
\end{figure}

As another typical type of engine-driven phenomena, many GRBs have
been suggested to harbor a magnetar engine too (Usov 1992; Duncan \&
Thompson 1992; Dai \& Lu 1998a,b; Wheeler et al. 2000; Zhang \&
Meszaros 2001; Thompson et al. 2004; Metzger et al. 2011), which is
strongly supported by the shallow-decay or plateau afterglows of
GRBs (Yu et al. 2010; Rowlinson et al. 2013; Yi et al. 2014; L{\"u}
\& Zhang 2014) and the rapidly rising and declining X-ray flares
(Burrows et al. 2005; Dai et al. 2006; Wang \& Dai 2013; Yi et al.
2016). In view of the same massive-star-collapse origin of long
GRBs, it is obviously of great importance to compare the magnetar
properties of SLSNe with those of long GRBs, which is one of the
primary purposes of this paper. Such a comparison is exhibited in
Figure \ref{figGRB}, where the parameters of GRB magnetars are taken
from the gold and silver samples in L\"u \& Zhang
(2014)\footnote{The derivation of the parameters of GRB magnetars is
dependent on the measurements of jet beaming angles, since the
energy release from a GRB magnetar is probably highly collimated
(see L\"u \& Zhang 2014 for detailed discussions). Moreover, the
assignment of the magnetar energy to GRB prompt emission, afterglow
emission, and associated hypernova emission also needs to be
analyzed carefully.}. As shown, on one hand, the initial spin
periods of GRB magnetars are generally concentrated within the range
from about 1 ms to 10 ms, which is similar to the case of SLSN
magnetars but has a relatively longer average value of $\langle
P_{\rm i}\rangle_{\rm GRB}=7.8$ ms. On the other hand, the magnetic
field strengths of GRB magnetars, of an average value of $\langle
B_{\rm p}\rangle_{\rm GRB}=6.3\times 10^{15}$G, are about several
tens of times higher than those of SLSN magnetars. The dividing line
between GRBs and SLSNe can roughly be set at a field strength of
$\sim4\times10^{14}$ G, which is about ten times of the lower
boundary of field strengths of SLSN magnetars. This significant
difference in magnetic fields may arise from intrinsically different
natures (e.g., neutron stars, hybrid stars, quark stars, etc.) of
the two types of magnetars. It is also suggested that ultrahigh
magnetic fields could play a crucial role in producing GRBs. Figure
\ref{figGRB} also reveals that the magnetic field strengths and
initial spin periods of the two types of magnetars could satisfy two
similar rough correlations, i.e., $B_{\rm p}\sim7.6\times
10^{13}P_{\rm i,-3}^{0.6}$ for ~SLSN~magnetars and $B_{\rm
p}\sim8.3\times 10^{14}P_{\rm i,-3}^{0.8}$ for ~GRB~magnetars. If
these correlations are true, then we could conclude that the
rotational energy of a newborn magnetar, no matter which kind of
material it consists of, is always approximately inversely
proportional to its magnetic energy, but with a large error. This
could provide a clue to explore the dynamo of magnetars.

\subsection{Explosion parameters}\label{SectMej}
\begin{figure}
\centering\resizebox{0.5\hsize}{!}{\includegraphics{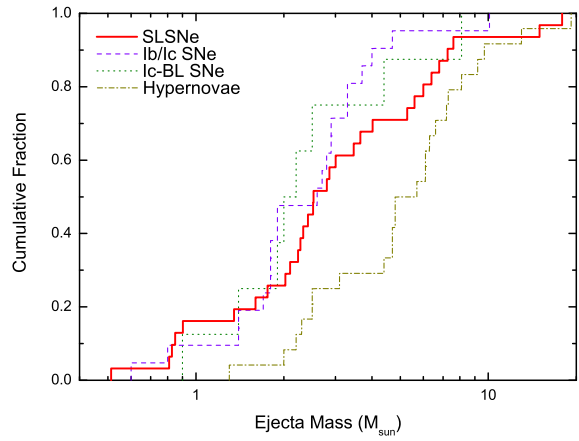}}
\caption{Accumulated distribution of ejecta masses of SLSNe (thick
solid line). The dashed, dotted, and dash-dotted lines correspond to
the cases of normal Ib/c, Ic-BL, and hypernovae,
respectively.}\label{figmej}
\end{figure}

The accumulated distribution of ejecta masses of the SLSNe is ploted
in Figure \ref{figmej}. For a comparison, the distributions of
GRB-associated hypernovae, normal Ic-BL supernovae that are
unassociated with GRBs, and normal Ib/c supernovae are also
presented, the data of which are taken from Cano et al. (2016) and
Lyman et al. (2014). As shown, the mass range of SLSN ejecta is
mainly from $\sim1M_{\odot}$ to $\sim10M_{\odot}$, which is roughly
consistent with the other types of supernovae. More specifically,
the SLSN distribution seems broadly intermediate between those of
normal supernovae and hypernovae, although larger numbers of the
comparative supernovae are still needed for a more reliable
comparison. The average ejecta masses of normal Ib/c supernovae,
Ic-BL supernovae, SLSNe, and hypernovae are $2.8M_{\odot}$,
$2.9M_{\odot}$, $4.0M_{\odot}$, and $6.0M_{\odot}$, respectively. On
one hand, as argued by Nicholl et al. (2015a), this result may
indicate that the ejecta masses could play the dominant role in
distinguishing SLSNe as well as hypernovae from normal supernovae.
On the other hand, in our opinion, these different mass
distributions could be a consequence/indication of different central
engines. One possibility is that the ejecta masses of SLSNe and
hypernovae were all overestimated, since an isotropic geometry was
always considered in calculations whereas the actual geometries of
the ejecta are probably very anisotropic.

\begin{figure}
\centering\resizebox{0.5\hsize}{!}{\includegraphics{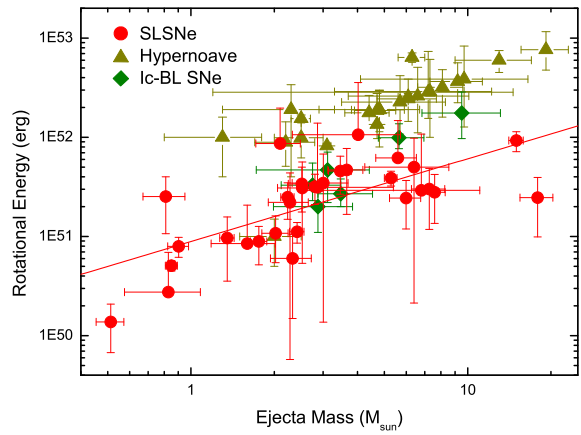}}
\caption{Relationship between ejected masses of SLSNe and rotational
energies of SLSN magnetars. The best-ftting log-linear relation
$E_{\rm rot}\propto M_{\rm ej}^{0.86}$ is showed by the solid line.
The masses and kinetic energies of the ejecta of Type Ic-BL
supernovae and hypernovae are showed for a
comparison. 
}\label{figmejErot}
\end{figure}

Figure \ref{figmej} also shows that the ejecta masses of about 16\%
SLSNe are less than $\sim1M_{\odot}$, which could indicate that
these SLSNe originate from extremely-stripped progenitors but still
might cause some concerns about their energy supplies and
velocities. Nevetheless, in the magnetar model, the radiated energy
of these SLSNe is provided by the magnetar engine, which is
therefore irrelevant to the ejecta mass. In fact, even for much
smaller ejecta of masses $\sim 10^{-3}-10^{-2}M_{\odot}$, some
analogous luminous transients have been suggested by Yu et al.
(2013, 2015) for the systems of double neutron star mergers or
accretion-induced collapses of white dwarfs. In these cases, the
ejecta could be accelerated to an extremely high velocity. Then,
would such high velocities also be predicted for low-mass SLSNe? In
order to answer this question, we display the ejecta masses of SLSNe
against the rotational energy of their remnant magnetars in Figure
\ref{figmejErot}, which exhibits a correlation as
\begin{eqnarray}
E_{\rm rot}={8.5\times10^{50}}(M_{\rm ej}/M_{\odot})^{0.86}\rm erg,\label{eqsErotMej}
\end{eqnarray}
where the rotational energy is calculated by $E_{\rm rot}=L_{\rm
sd,i}t_{\rm sd}$. By according to $v_{\rm f}\lesssim (2E_{\rm
rot}/M_{\rm ej})^{1/2}$, the final velocity of SLSN ejecta can be
found to be not much higher than $\sim10^{9}\rm cm~s^{-1}$, which is
consistent with SLSN observations. This result is insensitive to the
ejecta mass.

For a comparison, the masses and kinetic energies of the ejecta of
hypernovae as well as some normal Ic-BL supernovae are also
presented in Figure \ref{figmejErot}. We find with great interest
that these hypernovae/Ic-BL supernovae data could follow an $E_{\rm
kin}-M_{\rm ej}$  relationship similar to the SLSN $E_{\rm
rot}-M_{\rm ej}$ relationship. In our opinion, the ejecta kinetic
energy of a hypernova/Ic-BL supernova is ultimately provided by its
central engine. Here the ultrahigh kinetic energies of some
hypernovae on the order of $\sim10^{52}-10^{53}$ erg could be a
problem for the magnetar engine model. Nevertheless, as suspected
above, these kinetic energies could be somewhat overestimated due to
the neglecting of the high anisotropy of hypernovae. Furthermore, in
principle, an upper limit of several times $10^{52}$ erg on the
rotational energy of a magnetar could still be acceptable, by
considering of the possible differential rotation and uncertain
equations of state of the magnetar. After all, the physical nature
of the magnetars in hypernovae could be completely different from
SLSN magnetars, as inferred by the magnetic field statistics in
Section \ref{SectMag}. In summary, the consistency between the
possible $E_{\rm rot}(E_{\rm kin})-M_{\rm ej}$ relationships of
SLSNe and hypernovae/Ic-BL supernovae suggests that millisecond
magnetars could also be the central engines of hypernovae/Ic-BL
supernovae. However, different from SLSNe, the energy released from
the magnetars of hypernovae/Ic-BL supernovae is mainly used to
accelerate the ejecta rather than to power the supernova emission.

Finally, we would like to connect the $E_{\rm rot}-M_{\rm ej}$
relationship with the work of Chen et al. (2016a) who suggested a
possible correlation between the initial spin periods of SLSN
magnetars and the metallicities of their host galaxies, although,
which still needs to be examined by a much larger sample. In any
case, theoretically, for a lower metallicity, the pre-explosion mass
loss of a SLSN progenitor could be smaller and thus more angular
momentum can be hold by the progenitor. As a result, more mass can
be ejected during the supernova explosion and a more rapid rotating
magnetar can be formed because its angular momentum is completely
inherited from the core of the progenitor.

\subsection{Shapes of SLSN light curves}
The shapes of SLSN light curves are usually treated as a preliminary
basis to judge the possible properties of energy sources of the
SLSNe. For example, some slowly-decaying SLSNe were usually ascribed
to radioactive PISNe. In the framework of magnetar-powered
supernovae, an approximative analytical solution of Equations
(\ref{Eint}$-$\ref{Lp}) as presented in Yu et al. (2015) clearly
shows that the basic shape of a light curve is mainly determined by
the relationship between the spin-down timescale of the magnetar and
the heat diffusion timescale of the supernova ejecta. To be
specific, on one hand, a fast-evolving light curve can be obtained
if $t_{\rm sd}$ and $t_{\rm d}$ are both short. On the other hand,
for a relatively long $t_{\rm d}$ but $t_{\rm sd}\ll t_{\rm d}$, an
exponentially fast-fading light curve can still be obtained. In
other words, a fast-evolving light curve can always be produced with
a short spin-down timescale, which usually corresponds to a
relatively high magnetic field. Therefore, it can be expected that
the sharpness of SLSN light curves is tightly connected with the
magnetic fields of their central magnetars.

\begin{figure}
\centering\resizebox{0.5\hsize}{!}{\includegraphics{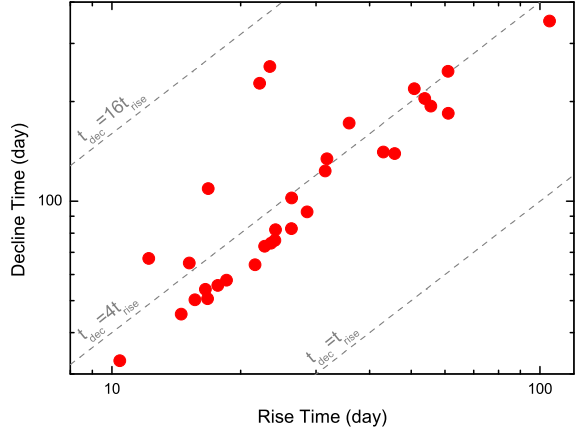}}
\caption{The rise timescales vs the decline timescales of SLSN light
curves. {\bf These timescales are derived from the best model fit of
the light curves}. The dashed lines represent different
relationships between these two timescales as
labeled.}\label{figtrd}
\end{figure}

For a direct description of light curve shapes, we define a rise
timescale $t_{\rm rise}$ and a decline timescale $t_{\rm dec}$,
during which SLSN luminosity respectively increases from $10\%$ of
peak value to the peak and decreases from the peak to its $10\%$. We
calculate these timescales with the model equations and the
observationally-constrained model parameters, as listed in Table
\ref{tablcpara}. Figure \ref{figtrd} shows that these two timescales
are well correlated with each other with a ratio around $t_{\rm
dec}/t_{\rm rise}\sim(3-4)$. Following Yu et al. (2015), the typical
value of this ratio can be calculated by
$(1/\sqrt{0.1}-1)/(1-\sqrt{0.1})=3.2$, by considering of the
simplest form of light curves in the magnetar engine model, i.e.,
$L_{\rm sn}\propto t^{2}$ for $t<t_{\rm peak}$ and $L_{\rm
sn}\propto t^{-2}$ for $t>t_{\rm peak}$, where $t_{\rm peak}$ is the
peak time. A similar result had been presented in Nicholl et al.
(2015a), where this $t_{\rm rise}-t_{\rm dec}$ correlation was
suggested to be an evidence for the magnetar engine model.

\begin{figure}
\centering\resizebox{0.5\hsize}{!}{\includegraphics{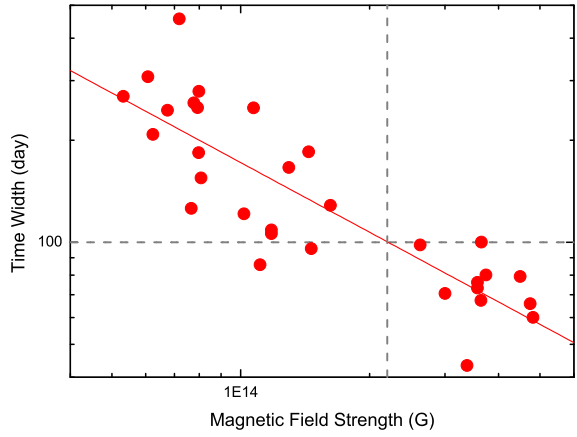}}
\caption{The peak widths of SLSN light curves vs the magnetic field
strengths of SLSN magentars. A possible correlation of $\Delta
t_{10\%}\propto B_{\rm p}^{-0.68}$ is represented by the solid line.
The dashed lines give the separating lines for the two subclasses of
SLSNe at the values of $B_{\rm p}=2.3\times10^{14}$ G and $\Delta
t_{10\%}=100$ day.}\label{figChi}
\end{figure}

\begin{figure}
\centering\resizebox{0.5\hsize}{!}{\includegraphics{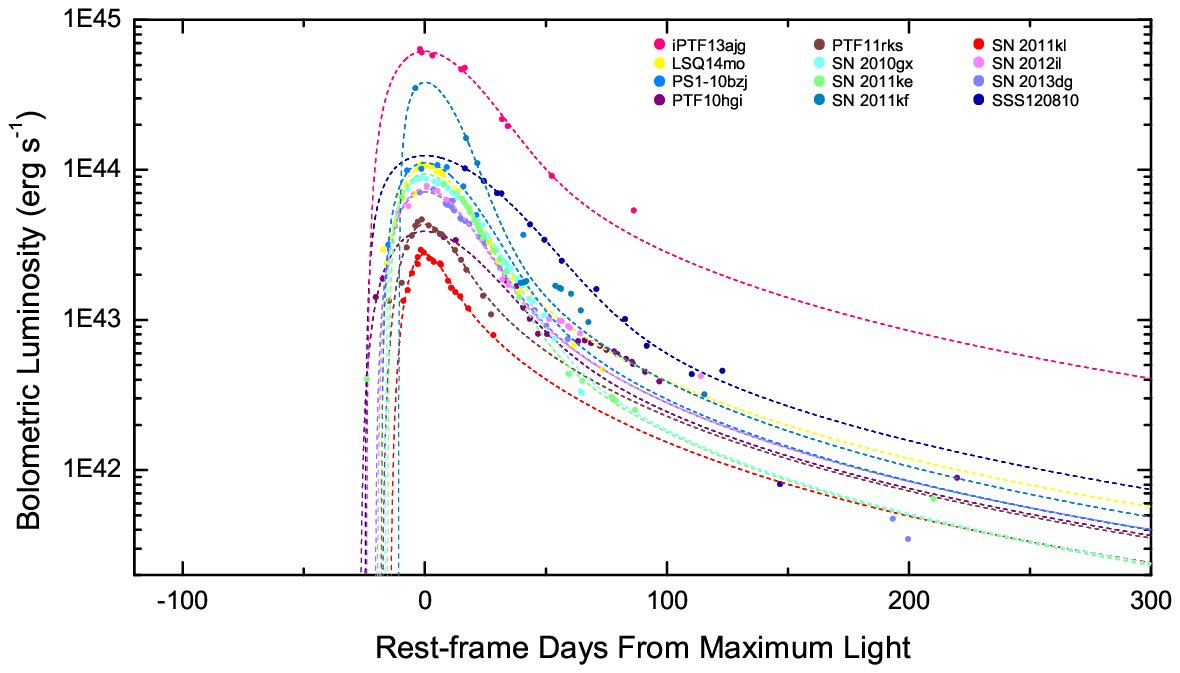}}\resizebox{0.5\hsize}{!}{\includegraphics{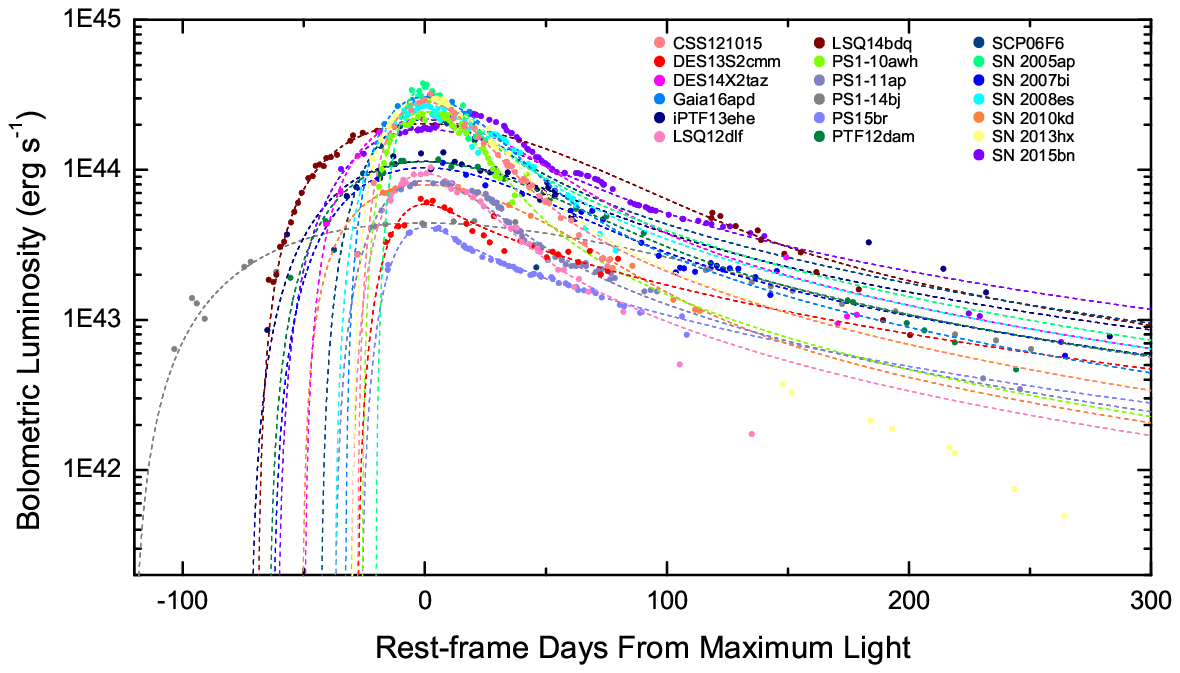}}
\caption{A collection of fast-evolving (left) and slow-evolving
(right) SLSN light curves.}\label{figLCs}
\end{figure}

The sum of $\Delta t_{10\%}=t_{\rm rise}+t_{\rm dec}$ can be used to
define the width of the peak of light curves and furthermore to
indicate the sharpness of light curves. As expected, Figure
\ref{figChi} indeed shows that the higher the magnetic fields, the
smaller the light curve widths, which can be roughly expressed as
$\Delta t_{10\%}\propto B_{\rm p}^{-0.68}$. Therefore, by
considering of the classification of SLSNe according to $B_{\rm p}$
as found in Section \ref{SectMag}, it is natural to expect that SLSN
light curves can also be divided into two types by a rough dividing
line of $\Delta t_{10\%}\sim 100$ day corresponding to $B_{\rm
p}\sim 2.2\times10^{14}$ G. Following such a criteria, we can
successfully classify the fast-evolving (high-field) and
slow-evolving (low-field) SLSN light curves into the left and right
panels of Figure \ref{figLCs}, respectively.


\section{Discussions on magnetar-driven explosions}\label{SectDis}
It has been widely discovered in observations that SLSNe have some
similarities and even intrinsic connections with long GRBs and their
associated hypernovae (van den Heuvel et al. 2013; Lunnan et al.
2015; Nicholl et al. 2015a; Yu \& Li 2016; Inserra et al. 2016b;
Japelj et al. 2016; Prajs et al. 2017). In particular, the host
galaxies of GRBs and SLSNe I are found to share many common
properties (e.g. high star formation rate and low metallicity;
Lunnan et al. 2014), although some (not surprising) differences
still exist (Angus et al. 2016). More directly, a SLSNe (i.e., SN
2011kl) was even discovered in the afterglow emission of GRB 111209A
(Levan et al. 2013; Greiner et al. 2015).

\begin{figure}
\centering\resizebox{0.9\hsize}{!}{\includegraphics{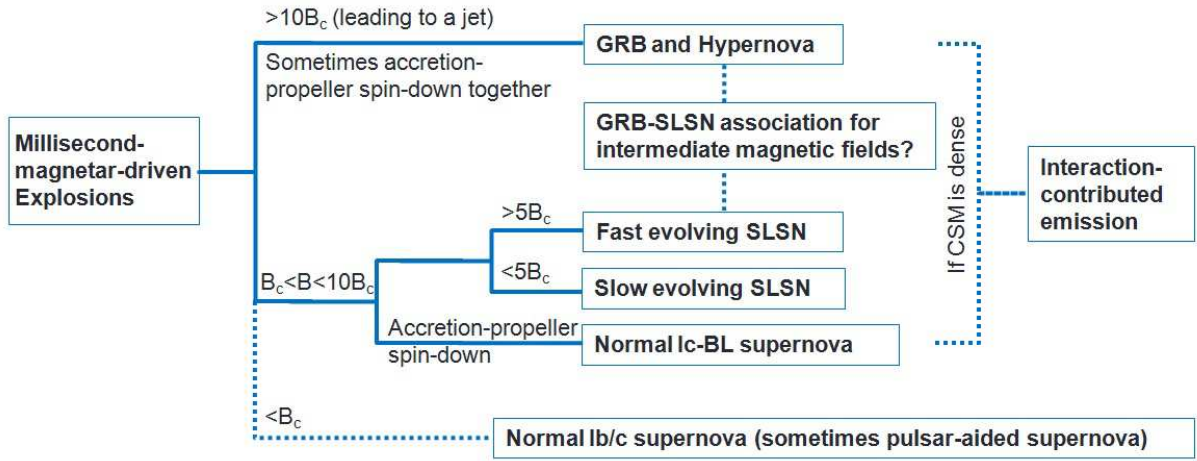}}
\caption{Possible connections between different magnetar-driven
explosion phenomena.}\label{cartoon}
\end{figure}

From a theoretical point of view, the most fundamental connection
between SLSNe and GRBs is that both of these two types of stellar
explosions could leave a remnant millisecond magnetar, although it
can not be ruled out that some other energy sources could be
responsible for a fraction of SLSNe (e.g., interaction-powered
SLSNe) and GRBs (e.g. black hole accretion-driven GRBs). Besides to
contribute a useful continuous energy supply,  a magnetar engine can
also be employed to effectively explain the intermittent flare
activities discovered in both GRB afterglow emission (Dai et al.
2006) and SLSN emission (Yu \& Li 2016). This united magnetar engine
model for both SLSNe and GRBs implies that these two explosion
phenomena probably have very similar progenitors, environments, and
even explosion processes. Meanwhile, differences between their
progenitors must also be remarkable, which is needed to explain the
completely different magnetic fields of their remnant magnetars. In
Figure \ref{cartoon}, we illustrate our preliminary consideration of
a possible physical classification of various stellar explosions in
the united magnetar engine model, as discussed as follows.

First of all, the magnetic fields of SLSN magnetars are nearly all
higher than a lower limit value that is around the Landau critical
field of electrons as $B_{\rm c}=m_{\rm
e}^2c^3/(q\hbar)=4.4\times10^{13}$ G, where $m_{\rm e}$ and $q$ are
the mass and charge of electrons and $\hbar$ is the reduced Planck
constant. Such a field boundary of clear physical meaning might not
just be a coincidence. For a magnetar of a supercritical magnetic
field, it might have some unique properties making the magnetar
physically different from normal pulsars, because the Landau
quantization of electrons can make the phase space of the electrons
very anisotropic. Therefore, we suspect that a supercritical
magnetic field could play a fundamental role in SLSN explosions,
besides to determine a spin-down timescale comparable to the heat
diffusion timescale of supernova ejecta.

While the magnetic fields of SLSN magnetars are mainly within the
range of $\sim (1-10)B_{\rm c}$, GRB magnetars are generally found
to own a field higher than $\sim 10B_{\rm c}$. In our opinion, these
completely different ranges of magnetic fields may be intrinsically
determined by the different interior ingredients of SLSN magnetars
and GRB magnetars, although their dynamo processes could still be
common as indicated by the similar rough $B_{\rm p}-P_{\rm i}$
relationships. The possible consequences of the ultrahigh magnetic
fields of GRB magnetars are of more interests here. For the field
range of $\sim (10-300)B_{\rm c}$, the number of electron Landau
levels is only several tens and even a few (e.g. Zheng \& Yu 2006),
in which cases electrons would behavior very close to a
one-dimension gas. As a result, some processes (e.g. neutrino
emission) of GRB magnetars must be dominated in the magnetic field
direction and thus the resulted hypernovae could be highly
anisotropic as considered in Section \ref{SectMej}.

The ultrahigh magnetic field of a GRB magnetar could play a crucial
role in the formation and collimation of a relativistic jet, by
coupling with a highly anisotropic neutrino wind during the first
$\sim10-100$ s (Metzger et al. 2011). During this phase, the
propeller effect of possible fallback accretion could also play a
part and extract the rotational energy of the magnetar.
Subsequently, during a relatively longer but still very short
timescale of $\sim 10^{2-4}$ s, the ultrahigh magnetic field can
further lead the magnetar to release its remaining rotational
energy, a faction of which can contribute to a plateau component to
the GRB afterglow emission (Yu et al. 2010). At the same time, a
comparable amount of energy could be injected into the supernova
ejecta in a timescale that is significantly shorter than the
diffusion timescale of months. As a result, the supernova ejecta
will be accelerated quickly and significantly and the resulted
hypernovae should own a broad-line featured spectrum. However, the
hypernova luminosity can not be effectively enhanced, because the
injected energy has been consumed much earlier than the supernova
emission. These characteristics have been widely confirmed by
hypernova observations (Galama et al. 1998; Hjorth et al. 2003;
Stanek et al. 2003). Furthermore, the kinetic energy of hypernovae
can basically reflect the total rotational energy of magnetars, as
we discussed on Figure \ref{figmejErot}.

While a magnetic field of $\sim(1-10)B_{\rm c}$ is benefit for
producing an unusual bright supernova emission, a field higher than
$\sim10B_{\rm c}$ could be indispensable for jet formation.
Therefore, it could be natural to conclude that GRB-SLSN
associations can only appear for magnetar engines of a magnetic
field around $\sim10B_{\rm c}$. It is found that the magnetar engine
of SN 2011kl indeed satisfies this condition. However, for this
case, the problem is that the magnetic field derived from the
afterglow of GRB 111209A is inconsistent with that derived from SN
2011kl (Kann et al. 2016). In any case, the present reported
SLSN-associated GRBs are all ultra-long GRBs, which might have a
progenitor completely different from those of typical long GRBs.
Then, it can not be ruled out that these SLSNe have a unique origin.
For example, Nakauchi et al. (2013) suggested these SLSNe to be
contributed by a jet cocoon that breaks out from a blue supergiant.

A remaining question is that what the nature of normal Ic-BL
supernovae, which are unassociated with GRBs. In view of their high
similarity with hypernovae, the most straightforward answer is to
connect them with magnetars of ultrahigh magnetic fields. However,
in this case, a further question would arise as that what suppresses
the formation of relativistic jets. Therefore, we prefer to suggest
that normal Ic-BL supernovae could be driven by magnetars similar to
SLSN ones. The difference is that fallback accretion could take
place for normal Ic-BL supernovae, as proposed by Piro \& Ott
(2011). The high magnetic field of a magnetar can lead the accretion
to quickly enter into a propeller phase. Then most rotational energy
of the magnetar is extracted by the propeller outflow during a very
short timescale. As a result, the supernova ejecta is accelerated
but the supernova emission is still powered by radioactivities,
which is similar to the hypernova cases.

As discussed, no matter which types of millisecond magnetars are
formed, the ejecta of all magnetar-driven supernovae including
SLSNe, hypernovae, and normal Ic-BL supernovae should eventually be
accelerated to a high speed. Therefore, as long as these explosions
happen in a dense environment, a strong shock emission can always
arise and provide a substantial and even dominative contribution to
the supernova emission, i.e., an interaction-powered SLSN could be
generated. For example, an interaction signature of narrow lines
could finally appear at a few hundreds of days after peak emission
for some broad-lined SLSNe II (e.g. 2013hx). It is further suspected
that, even for SLSN IIn, their strong interactions could also
ultimately be resulted from the powering of a magnetar engine. The
intrinsic differences between SLSNe I and II could be not primarily
determined by their energy engines, but be a reflection of their
different environments and host galaxies.

Finally, in view of the wide and significant influences of
millisecond magnetars on the various unusual supernova phenomena, it
can be enlightened that some ordinary supernova emission could also
be partially powered or aided by a pulsar engine, even though the
magnetic field of the pulsar is normal (e.g. Li et al. 2015 for SN
1054). This idea could be supported by the wide existences of
engine-driven jets in the remnants of many ordianry core-collapse
supernovae (Bear \& Soker 2016a,b).

\section{Summary and conclusion}
By considering of that the majority of observed SLSNe are powered by
continuous energy injection from a magnetar engine, we fit the
bolometric light curves of 27 SLSNe I and 4 SLSNe II without
narrow-line features. As a result, we obtain the basic parameters of
these SLSNe including their ejecta masses and the initial spin
periods and magnetic field strengths of remnant magnetars. The
primary ranges of these parameters are found to be $\sim
(1-10)M_{\odot}$, $\sim(1-10)$ ms, and $\sim (1-10)B_{\rm c}$,
respectively, with average values of $4.0M_{\odot}$,  $3.5$ms, and
$2.0\times10^{14}$ G, by assuming $\kappa=0.1\rm cm^2g^{-1}$. In
more details, SLSN magnetars can be divided into two subclasses by a
magnetic field strength of $\sim2.2\times10^{14}$ G. Furthermore,
the bolometric light curves being fast-evolving and slow-evolving in
observations are found to be tightly connected with these high- and
low-field subclasses of magnetars, respectively.

By connecting with long GRBs and their associated hypernovae, it is
suggested that high magnetic fields of magnetars could play a
crucial role in distinguishing hypernovae and SLSNe from normal
supernovae and even in distinguishing themselves. As a possible
consequence, the high magnetic fields could cause an anisotropic
geometry of the supernovae and even lead to the formation of a
highly beamed jet. Therefore, on one hand, it is of great importance
to investigate in future the possible influences of high magnetic
fields on the behaviors of newly-born millisecond magnetars, the
knowledge of which is very limited now. On the other hand, it is
also important to explore the physical reasons being responsible for
the formation of the high magnetic fields. Fall-back accretions with
different accretion rates could play a part in these processes.

Besides the common magnetar engines, the connection between SLSNe
and long GRBs could also be exhibited in the distributions of their
ejecta masses. In particular, the possible correlations between
magnetar properties and ejecta masses as well as metallicities of
host galaxies (Chen et al. 2016a) could provide a very important
clue to explore the nature of the progenitors of these stellar
explosions.

\acknowledgements The authors acknowledge Z. G. Dai for his useful
discussions. This work is supported by the National Natural Science
Foundation of China (grant No. 11473008) and the Program for New
Century Excellent Talents in University (grant No. NCET-13-0822).

\begin{table*}
\centering \caption{Observational Information of SLSNe}\label{tabobspara}
\renewcommand{\arraystretch}{2.0}
\begin{tabular}{cccccc}
 \hline
 \hline
 SLSNe        &  Types      &      R.A.                           &      Decl.                       &   $z$      & References\\
\hline
 CSS121015    &  II   & $00^{h}42^{m}44^{s}.38$             & $+13^{\circ}28{'}26{''}.197$     &   $0.2868$ & Benetti+2014 \\
 DES13S2cmm   &  I        & $02^{h}42^{m}32^{s}.82$             & $-01^{\circ}21{'}30{''}.1$       &   $0.663$  & Papadopoulos+2015  \\
 DES14X3taz   &  I        & $02^{h}28^{m}04^{s}.46$             & $-04^{\circ}05{'}12{''}.7$       &   $0.608$  & Smith+2016 \\
 Gaia16apd    &  I       & $12^{h}02^{m}51^{s}.71$             & $+44^{\circ}15{'}27{''}.40$      &   $0.1018$ & Yan+2016; Nicholl+2016\\
 iPTF13ajg    &  I        & $16^{h}39^{m}03^{s}.95$             & $+37^{\circ}01{'}38{''}.4$       &   $0.7403$ & Vreeswijk+2014 \\
 iPTF13ehe    &  I        & $06^{h}53^{m}21^{s}.50$             & $+67^{\circ}07{'}56{''}.0$       &   $0.3434$ & Yan+2015; Wang+2016 \\
 LSQ12dlf     &  I       & $01^{h}50^{m}29^{s}.8$              & $-21^{\circ}48{'}45{''}.4$       &   $0.255$  & Nicholl+2014 \\
 LSQ14bdq     &  I       & $10^{h}01^{m}41^{s}.60$             & $-12^{\circ}22{'}13{''}.4$       &   $0.345$  & Nicholl+2015 \\
 LSQ14mo      &  I       & $10^{h}22^{m}41^{s}.53$             & $-16^{\circ}55{'}14{''}.4$       &   $0.2563$ & Chen+2016\\
 PS1$-$10awh  &  I        & $22^{h}14^{m}29^{s}.831$            & $-00^{\circ}04{'}03{''}.62$      &   $0.908$ & Chomiuk+2011 \\
 PS1$-$10bzj  &  I        & $03^{h}31^{m}39^{s}.862$            & $-27^{\circ}47{'}42{''}.17$      &   $0.650$  & Lunnan+2013    \\
 PS1$-$11ap   &  I       & $10^{h}48^{m}27^{s}.73$             & $+57^{\circ}09{'}09{''}.2$       &   $0.524$  & McCrum+2014 \\
 PS1$-$14bj   &  I        & $10^{h}02^{m}08^{s}.433$            & $+03^{\circ}39{'}19{''}.02$      &   $0.5215$ & Lunnan+2016 \\
 PS15br       &  II      & $11^{h}25^{m}19^{s}.22$             & $+08^{\circ}14{'}18{''}.9$       &   $0.101$  & Inserra+2016    \\
 PTF10hgi     &  I       & $16^{h}37^{m}47^{s}.08$             & $+06^{\circ}12{'}32{''}.35$      &   $0.100$  & Inserra+2013 \\
 PTF11rks     &  I       & $01^{h}39^{m}45^{s}.49$             & $+29^{\circ}55{'}26{''}.87$      &   $0.190$  & Inserra+2013 \\
 PTF12dam     &  I       & $14^{h}24^{m}46^{s}.20$             & $+46^{\circ}13{'}48{''}.3$       &   $0.107$  & Chen+2015 \\
 SCP06F6      &  I        & $14^{h}32^{m}27^{s}.395$            & $+33^{\circ}32{'}24{''}.83$      &   $1.189$  &  Berkeley+2009; Quimby+2011; Chatzopoulos+2013 \\
 SN 2005ap    &  I        & $13^{h}01^{m}14^{s}.83$             & $+27^{\circ}43{'}32{''}.3$       &   $0.2832$ & Quimby+2007; Chatzopoulos+2013 \\
 SN 2007bi    &  I       & $13^{h}19^{m}20^{s}.2$              & $+08^{\circ}55{'}44{''}.0$       &   $0.1279$ & Gal-Yam+2009; Chatzopoulos+2013 \\
 SN 2008es    &  II      & $11^{h}56^{m}49^{s}.06$             & $+54^{\circ}27{'}25{''}.77$      &   $0.213$  & Miller+2009; Gezari+2009; Chatzopoulos+2013 \\
 SN 2010gx    &  I       & $11^{h}25^{m}46^{s}.71$             & $-08^{\circ}49{'}41{''}.4$       &   $0.23$   & Pastorello+2010 \\
 SN 2010kd    &  I        & $12^{h}08^{m}01^{s}.11$             & $+49^{\circ}13{'}31{''}.1$       &   $0.101$  & Chatzopoulos+2013 \\
 SN 2011ke    &  I       & $13^{h}50^{m}57^{s}.78$             & $+26^{\circ}16{'}42{''}.40$      &   $0.143$  & Inserra+2013 \\
 SN 2011kf    &  I       & $14^{h}36^{m}57^{s}.53$             & $+16^{\circ}30{'}56{''}.6$       &   $0.245$  & Drake+2012; Inserra+2013 \\
 SN 2011kl    &  I       & $00^{h}57^{m}22^{s}.64$             & $-46^{\circ}48{'}03{''}.6$       &   $0.677$  & Greiner+2015\\
 SN 2012il    &  I       & $09^{h}46^{m}12^{s}.91$             & $+19^{\circ}50{'}28{''}.70$      &   $0.175$  & Inserra+2013 \\
 SN 2013dg    &  I       & $13^{h}18^{m}41^{s}.38$             & $-07^{\circ}04{'}43{''}.1$       &   $0.265$  & Nicholl+2014 \\
 SN 2013hx    &  II  & $01^{h}35^{m}32^{s}.83$             & $-57^{\circ}57{'}50{''}.6$       &   $0.125$  & Inserra+2016   \\
 SN 2015bn    &  I        & $11^{h}33^{m}41^{s}.55$             & $+00^{\circ}43{'}33{''}.4$       &   $0.1136$ & Nicholl+2016 \\
 SSS120810    &  I       & $23^{h}18^{m}01^{s}.8$              & $-56^{\circ}09{'}25{''}.6$       &   $0.156$  & Nicholl+2014 \\

\hline \hline
\end{tabular}
\end{table*}

\begin{table}
\centering \caption{The Values of Model Parameters}\label{tabfitpara}
\renewcommand{\arraystretch}{2.0}
\begin{tabular}{ccccccccc}
 \hline
 \hline
SLSN                & $M_{\rm ej}/M_{\odot}$   & $L_{\rm sd,i}/\rm 10^{45}erg~s^{-1}$       & $t_{\rm sd}/$day   &$B_{\rm p}/10^{13}$G       &$P_{\rm i}/$ms   & $E_{\rm rot}/10^{50}$erg   \\
\hline
 CSS121015          & $2.51\pm 0.27$  & $3.99\pm 1.12$  & $9.82\pm 1.94$      & $11.80\pm 2.33$    & $2.43\pm 0.58$      & $33.88\pm 16.18$      \\
 DES13S2cmm         & $0.90\pm 0.08$  & $0.10\pm 0.01$  & $91.03\pm 12.56$    & $7.99\pm 1.10$     & $5.01\pm 0.57$      & $7.97\pm 1.82$        \\
 DES14X3taz         & $6.40\pm 2.15$  & $4.00\pm 2.47$  & $14.53\pm 4.93$     & $7.97\pm 2.71$     & $2.00\pm 0.96$      & $50.16\pm 48.02$       \\
 Gaia16apd          & $3.65\pm 0.53$  & $5.68\pm 2.19$  & $9.55\pm 2.47$      & $10.17\pm 2.63$    & $2.07\pm 0.66$      & $46.88\pm 30.18$       \\
 iPTF13ajp          & $4.02\pm 1.70$  & $34.60\pm 50.51$& $3.55\pm 3.23$      & $11.10\pm 10.11$   & $1.37\pm 1.63$      & $106.0\pm 251.3$       \\
 iPTF13ehe          & $6.79\pm 4.26$  & $0.79\pm 1.16$  & $43.02\pm 52.01$    & $6.07\pm 7.34$     & $2.62\pm 3.52$      & $29.18\pm 78.39$       \\
 LSQ12dlf           & $2.02\pm 0.22$  & $0.76\pm 0.21$  & $16.38\pm 3.52$     & $16.20\pm 3.48$    & $4.31\pm 1.07$      & $10.77\pm 5.33$        \\
 LSQ14bdq           & $14.98\pm 0.90$ & $9.75\pm 1.38$  & $11.00\pm 0.96$     & $6.74\pm 0.59$     & $1.47\pm 0.17$      & $92.63\pm 21.15$       \\
 LSQ14mo            & $1.35\pm 0.08$  & $2.09\pm 0.41$  & $5.34\pm 2.33$      & $29.98\pm 13.10$   & $4.55\pm 1.44$      & $9.64\pm 6.09$         \\
 PS1$-$10awh        & $2.52\pm 0.47$  & $5.08\pm 2.53$  & $7.03\pm 2.31$     & $14.60\pm 4.80$    & $2.55\pm 1.05$      & $30.86\pm 25.86$          \\
 PS1$-$10bzj        & $2.28\pm 0.38$  & $16.23\pm 1.00$ & $1.58\pm 0.57$      & $36.39\pm 13.02$   & $3.01\pm 1.46$      & $22.15\pm 21.57$         \\
 PS1$-$11ap         & $2.42\pm 0.15$  & $0.52\pm 0.07$  & $24.72\pm 2.58$     & $12.94\pm 0.14$    & $4.23\pm 0.50$      & $11.18\pm 2.63$        \\
 PS1$-$14bj         & $17.89\pm 2.43$ & $0.79\pm 0.28$  & $36.32\pm 8.73$     & $7.19\pm 1.73$     & $2.85\pm 0.85$      & $24.67\pm 14.85$       \\
 PS15br             & $0.85\pm 0.04$  & $0.07\pm 0.00$  & $79.38\pm 5.43$     & $10.71\pm 0.73$    & $6.27\pm 0.37$      & $5.09\pm 0.60$         \\
 PTF10hgi           & $2.33\pm 0.25$  & $1.20\pm 0.57$  & $5.80\pm 1.63$      & $36.45\pm 10.23$   & $5.77\pm 2.17$      & $6.01\pm 4.52$         \\
 PTF11rks           & $0.83\pm 0.25$  & $0.26\pm 0.22$  & $12.06\pm 8.32$     & $37.37\pm 25.76$   & $8.53\pm 6.45$      & $2.75\pm 4.16$          \\
 PTF12dam           & $7.26\pm 1.00$  & $1.43\pm 0.51$  & $24.41\pm 6.18$     & $7.93\pm 2.01$     & $2.57\pm 0.78$      & $30.17\pm 18.39$        \\
 SCP06F6            & $2.86\pm 2.53$  & $0.95\pm 1.74$  & $36.91\pm 61.10$    & $6.23\pm 10.01$    & $2.53\pm 4.34$      & $31.30\pm 107.5$       \\
 SN 2005ap          & $0.81\pm 0.14$  & $0.94\pm 0.25$  & $31.44\pm 6.87$     & $7.66\pm 2.47$     & $2.81\pm 0.81$      & $25.30\pm 14.64$        \\
 SN 2007bi          & $5.99\pm 0.77$  & $0.90\pm 0.25$  & $31.44\pm 6.87$     & $7.77\pm 1.70$     & $2.87\pm 0.73$      & $24.36\pm 12.43$      \\
 SN 2008es          & $3.01\pm 0.69$  & $1.95\pm 1.05$  & $20.51\pm 8.71$     & $8.08\pm 3.43$     & $2.40\pm 1.15$      & $34.62\pm 33.25$       \\
 SN 2010gx          & $2.80\pm 0.12$  & $56.61\pm 12.15$& $0.65\pm 0.08$      & $47.39\pm 5.80$    & $2.51\pm 0.42$      & $31.75\pm 10.70$      \\
 SN 2010kd          & $7.60\pm 0.69$  & $4.04\pm 1.28$  & $7.99\pm 1.57$      & $14.41\pm 2.83$    & $2.68\pm 0.69$      & $27.91\pm 14.32$      \\
 SN 2011ke          & $2.23\pm 0.13$  & $35.90\pm 9.09$ & $0.80\pm 0.12$      & $48.15\pm 7.11$    & $2.83\pm 0.57$      & $24.89\pm 9.98$       \\
 SN 2011kf          & $2.10\pm 0.38$  & $2141\pm 1723$  & $0.47\pm 0.22$      & $33.77\pm 15.86$   & $1.52\pm 0.97$      & $86.65\pm 110.5$        \\
 SN 2011kl          & $0.51\pm 0.06$  & $0.10\pm 0.02$  & $16.63\pm 4.31$     & $44.94\pm 1.17$    & $12.04\pm 3.07$     & $1.38\pm 0.70$        \\
 SN 2012il          & $1.60\pm 0.42$  & $2.29\pm 2.02$  & $4.29\pm 2.39$      & $35.71\pm 19.87$   & $4.86\pm 3.50$      & $8.47\pm 12.20$       \\
 SN 2013dg          & $1.76\pm 0.12$  & $2.53\pm 0.64$  & $4.08\pm 0.67$      & $35.69\pm 5.83$    & $4.74\pm 0.99$      & $8.91\pm 3.73$        \\
 SN 2013hx          & $3.47\pm 0.33$  & $7.24\pm 1.82$  & $7.29\pm 1.20$      & $11.79\pm 1.94$    & $2.09\pm 0.43$      & $45.64\pm 18.96$      \\
 SN 2015bn          & $5.28\pm 0.28$  & $1.07\pm 0.10$  & $42.09\pm 3.11$     & $5.32\pm 0.39$     & $2.27\pm 0.19$      & $38.90\pm 6.55$       \\
 SSS120810          & $5.59\pm 0.94$  & $66.09\pm 58.26$& $1.09\pm 0.55$      & $26.24\pm 13.27$   & $1.80\pm 1.25$      & $61.97\pm 85.96$       \\

\hline \hline
\end{tabular}
\end{table}

\begin{table}
\centering \caption{The Light Curve Parameters}\label{tablcpara}
\renewcommand{\arraystretch}{2.0}
\begin{tabular}{ccccccccc}
 \hline
 \hline
SLSN                & $\Delta t_{10\%}$  & $t_{\rm rise}$ &  $t_{\rm dec}$  \\
\hline
 CSS121015          &   $106.1$     &    $24.12$    &    $81.93$    &  \\
 DES13S2cmm         &   $278.7$     &    $23.40$    &    $255.3$    &  \\
 DES14X3taz         &   $183.8$     &    $43.08$    &    $140.7$    &\\
 Gaia16apd          &   $121.4$     &    $28.58$    &    $92.82$    &  \\
 iPTF13ajp          &   $85.82$     &    $21.60$    &    $64.22$    &  \\
 iPTF13ehe          &   $308.1$     &    $60.97$    &    $247.1$    & \\
 LSQ12dlf           &   $128.6$     &    $26.28$    &    $102.3$    &   \\
 LSQ14bdq           &   $245.4$     &    $61.04$    &    $184.4$    & \\
 LSQ14mo            &   $70.66$     &    $16.53$    &    $54.12$    &  \\
 PS1$-$10awh         &   $95.79$     &    $22.75$    &    $73.03$    &   \\
 PS1$-$10bzj        &   $67.42$     &    $16.74$    &    $50.68$    &   \\
 PS1$-$11ap         &   $166.2$     &    $31.81$    &    $134.4$    &   \\
 PS1$-$14bj         &   $456.1$     &    $105.3$    &    $350.8$    &    \\
 PS15br             &   $249.3$     &    $22.13$    &    $227.2$    &    \\
 PTF10hgi           &   $100.1$     &    $24.02$    &    $76.06$    &    \\
 PTF11rks           &   $80.12$     &    $15.17$    &    $64.94$    &    \\
 PTF12dam           &   $249.5$     &    $55.58$    &    $194.0$    &   \\
 SCP06F6            &   $208.0$     &    $35.83$    &    $172.2$    &   \\
 SN 2005ap          &   $126.0$     &    $16.78$    &    $109.2$    &   \\
 SN 2007bi          &   $258.1$     &    $53.77$    &    $204.3$    &   \\
 SN 2008es          &   $154.8$     &    $31.50$    &    $123.3$    &   \\
 SN 2010gx          &   $65.06$     &    $15.64$    &    $50.32$    &   \\
 SN 2010kd          &   $185.0$     &    $45.76$    &    $139.3$    &  \\
 SN 2011ke          &   $60.06$     &    $14.51$    &    $45.54$    &   \\
 SN 2011kf          &   $43.34$     &    $10.44$    &    $32.90$    &   \\
 SN 2011kl          &   $79.28$     &    $12.20$    &    $67.08$    &   \\
 SN 2012il          &   $73.31$     &    $17.69$    &    $55.63$    &   \\
 SN 2013dg          &   $76.16$     &    $18.52$    &    $57.64$    &   \\
 SN 2013hx          &   $108.9$     &    $26.26$    &    $82.59$    &   \\
 SN 2015bn          &   $269.7$     &    $50.87$    &    $218.8$    &  \\
 SSS120810          &   $98.17$     &    $23.52$    &    $74.65$    &   \\

\hline \hline
\end{tabular}
\end{table}

\begin{figure}
\centering\resizebox{0.3\hsize}{!}{\includegraphics{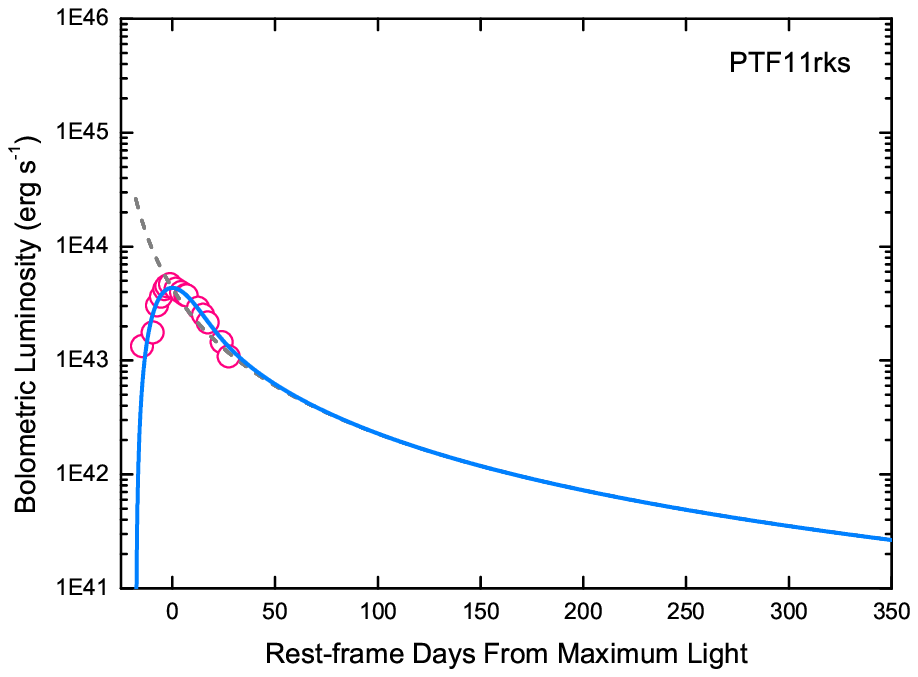}}\resizebox{0.3\hsize}{!}{\includegraphics{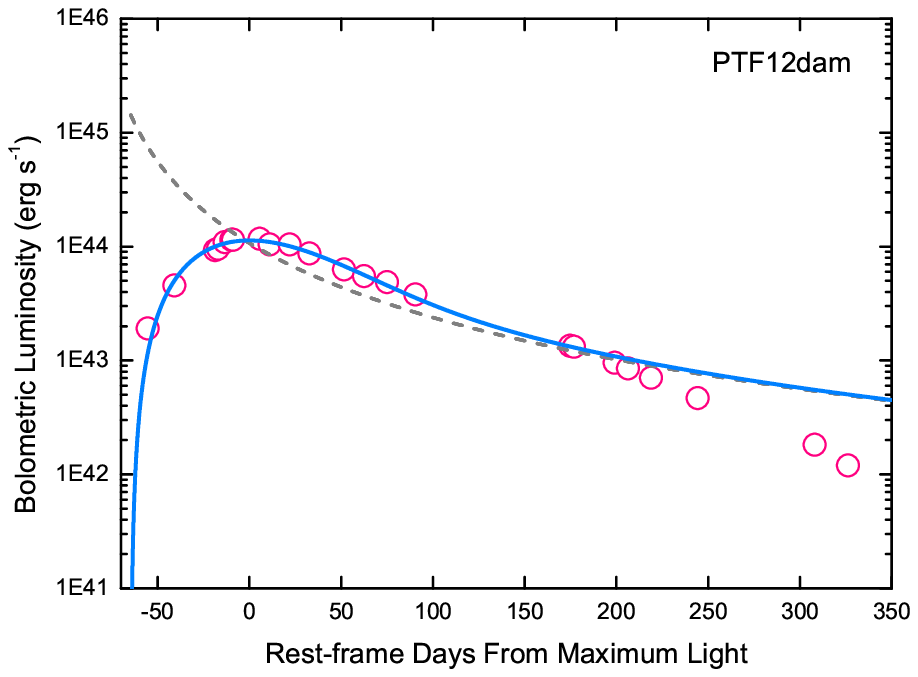}}\resizebox{0.3\hsize}{!}{\includegraphics{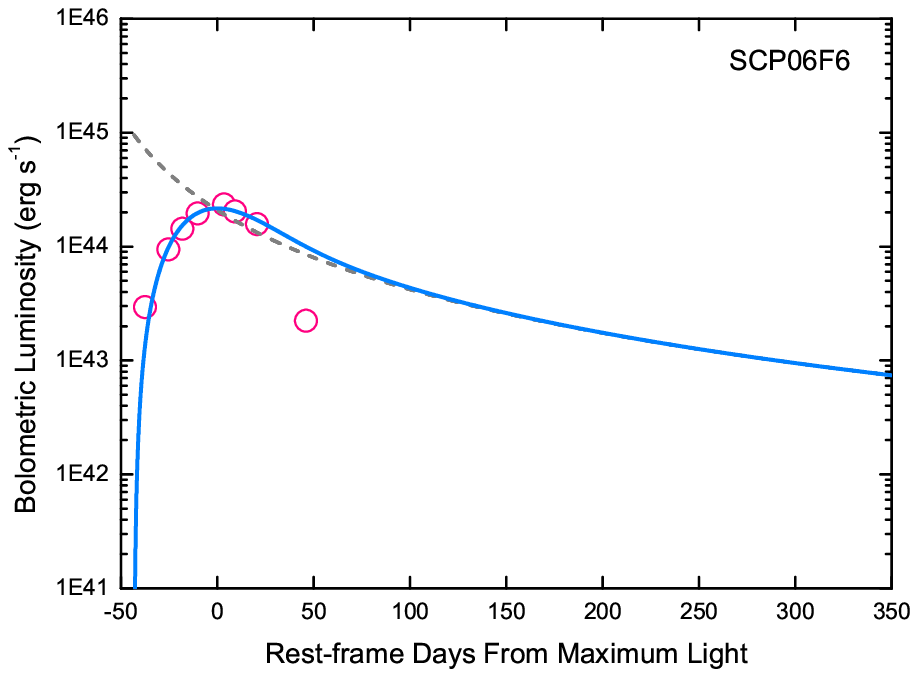}}
\centering\resizebox{0.3\hsize}{!}{\includegraphics{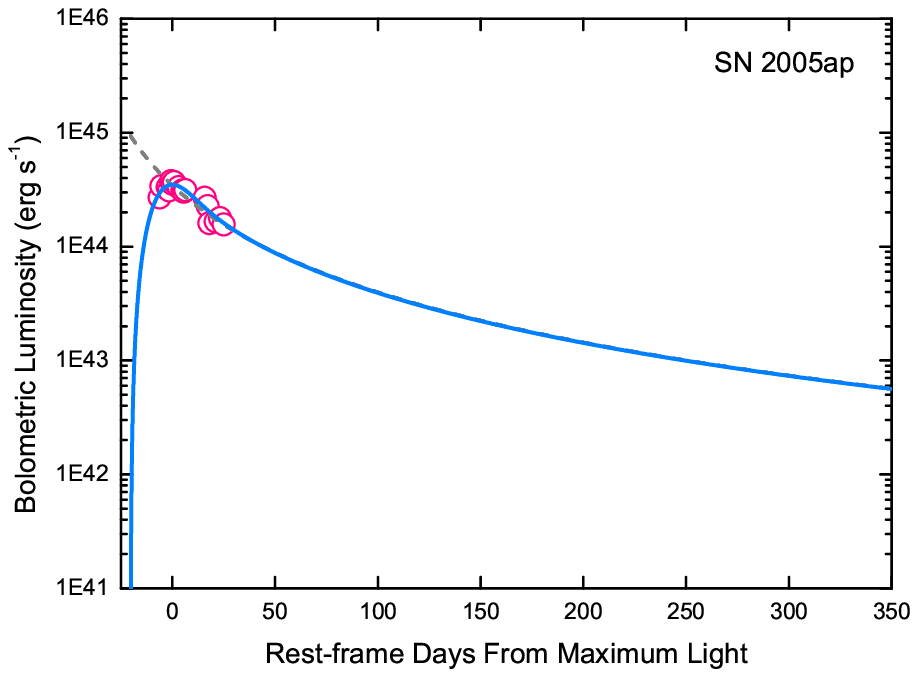}}\resizebox{0.3\hsize}{!}{\includegraphics{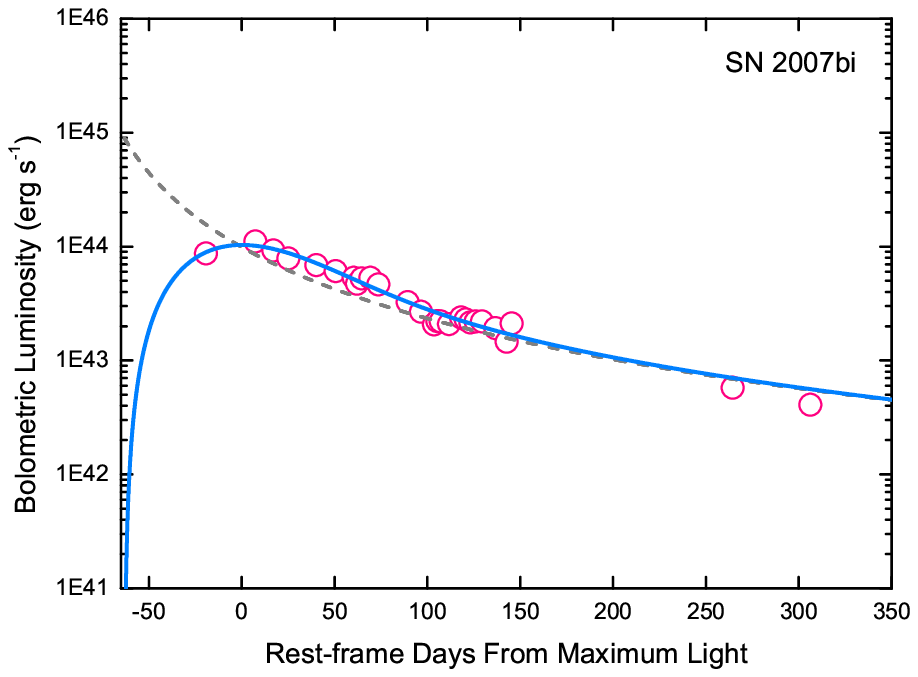}}\resizebox{0.3\hsize}{!}{\includegraphics{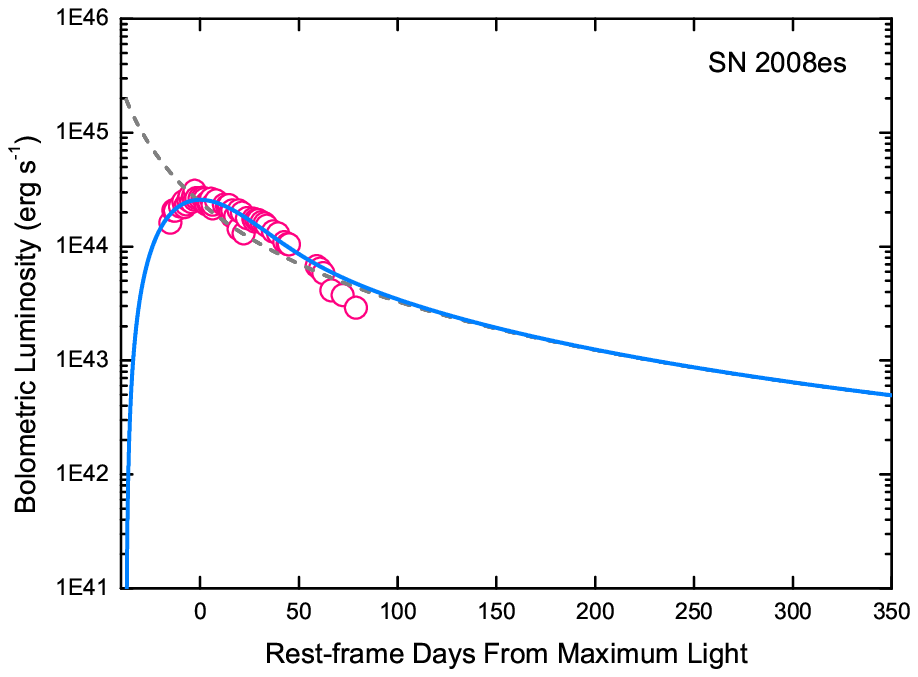}}
\centering\resizebox{0.3\hsize}{!}{\includegraphics{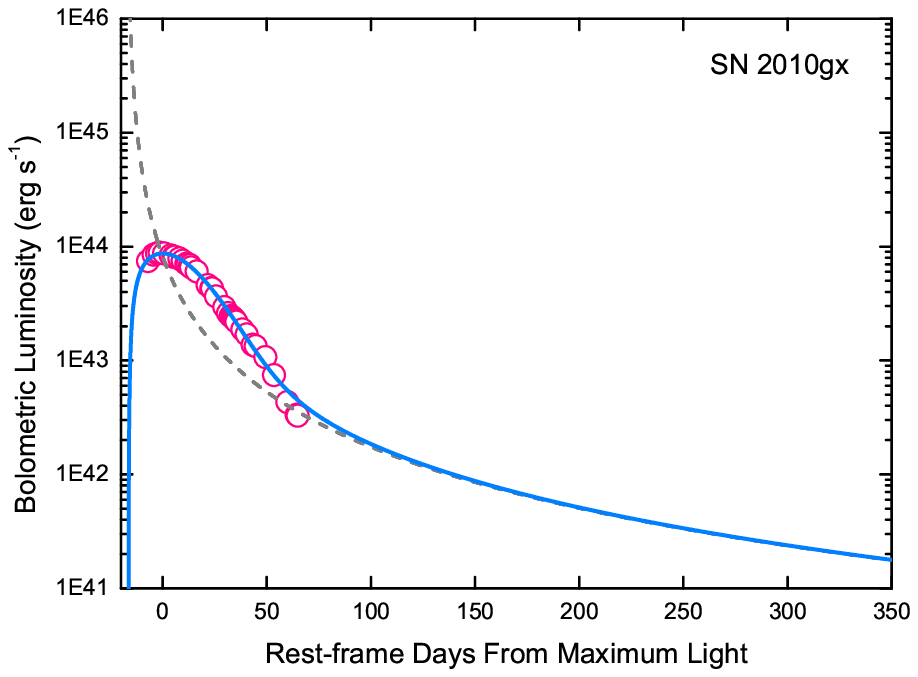}}\resizebox{0.3\hsize}{!}{\includegraphics{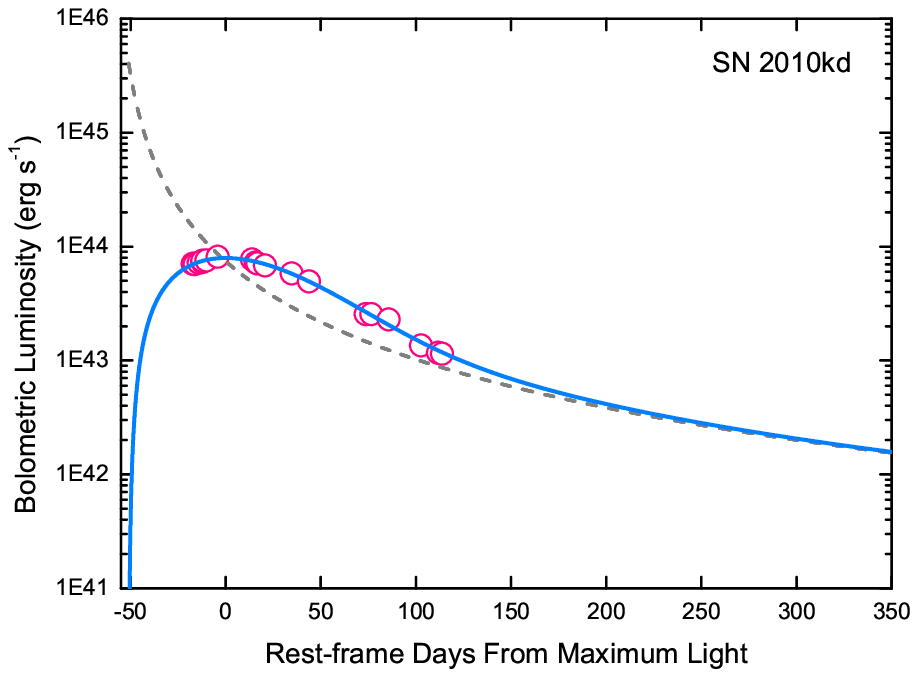}}\resizebox{0.3\hsize}{!}{\includegraphics{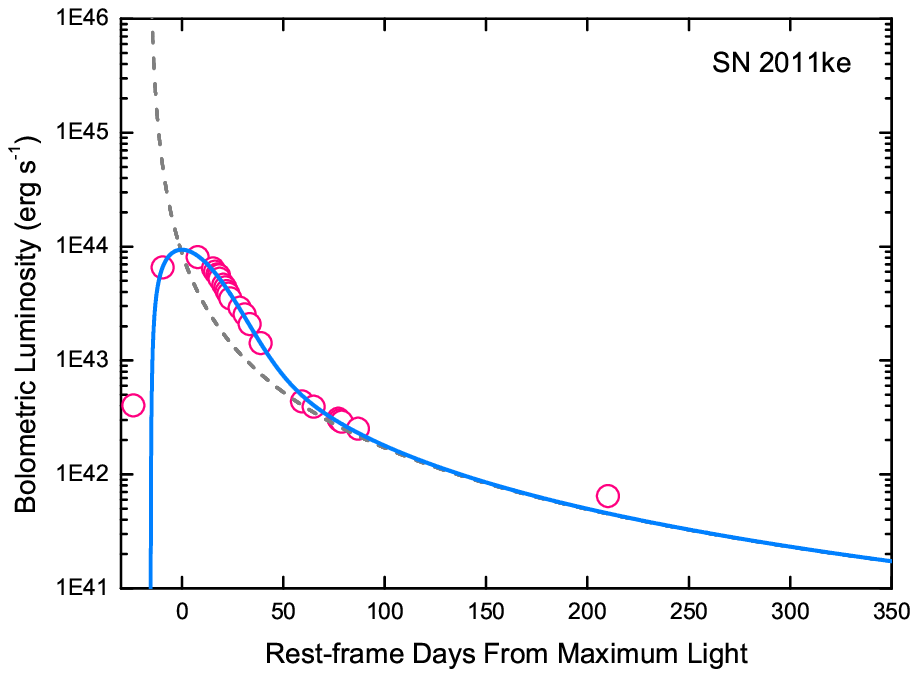}}
\centering\resizebox{0.3\hsize}{!}{\includegraphics{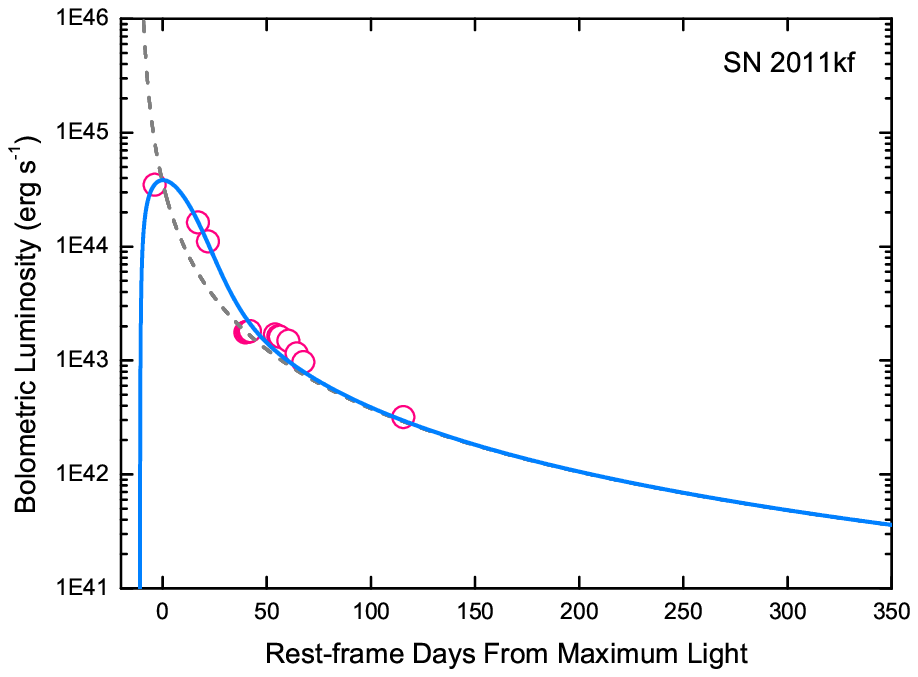}}\resizebox{0.3\hsize}{!}{\includegraphics{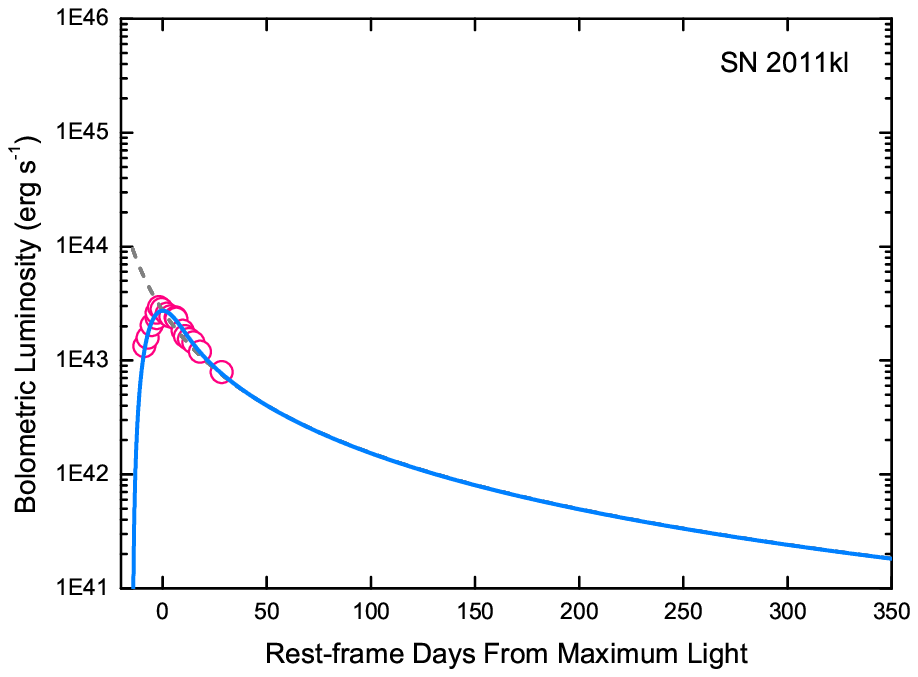}}\resizebox{0.3\hsize}{!}{\includegraphics{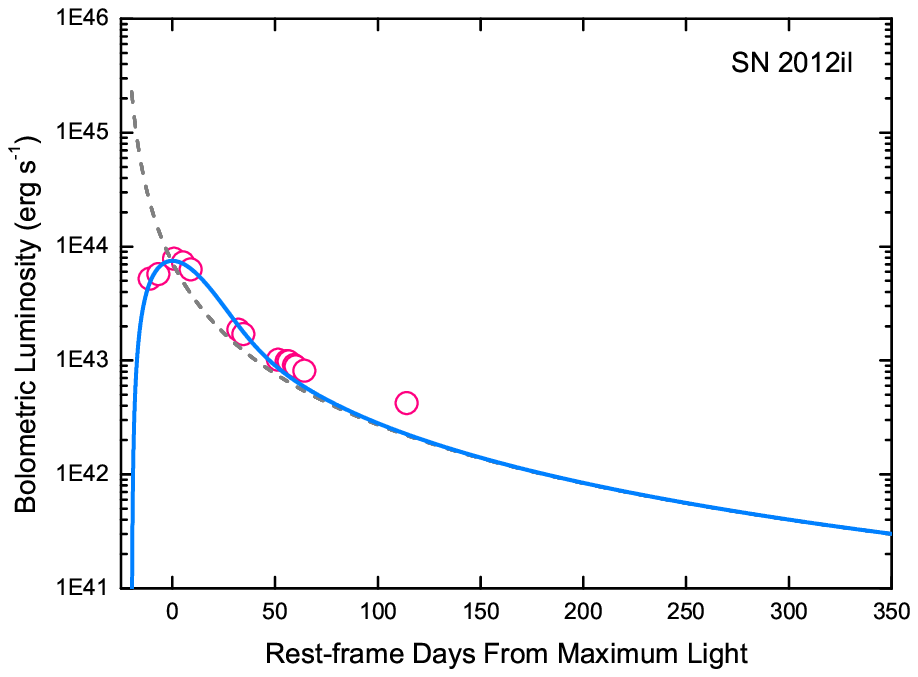}}
\centering\resizebox{0.3\hsize}{!}{\includegraphics{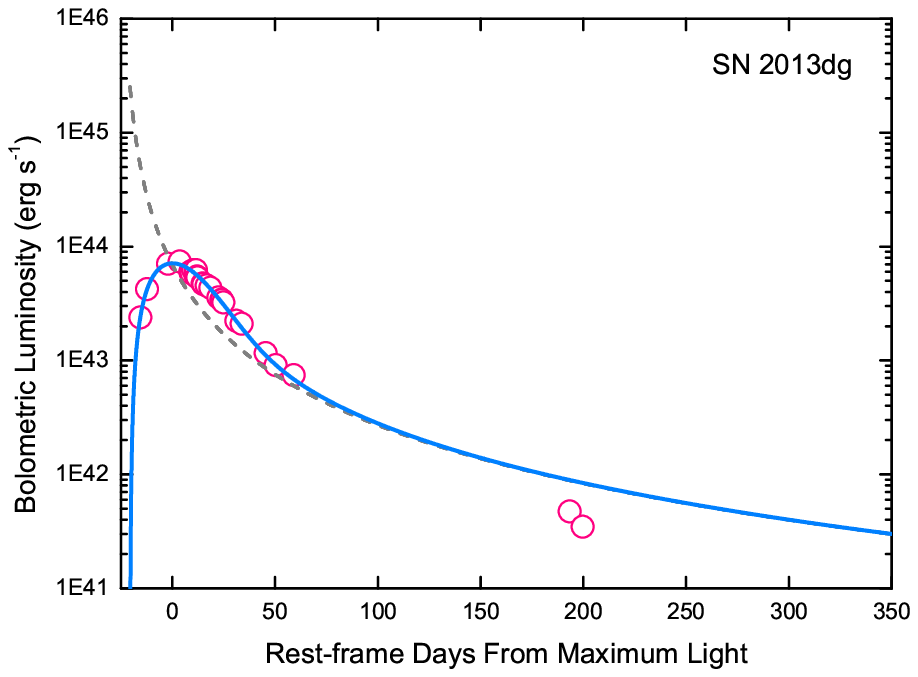}}\resizebox{0.3\hsize}{!}{\includegraphics{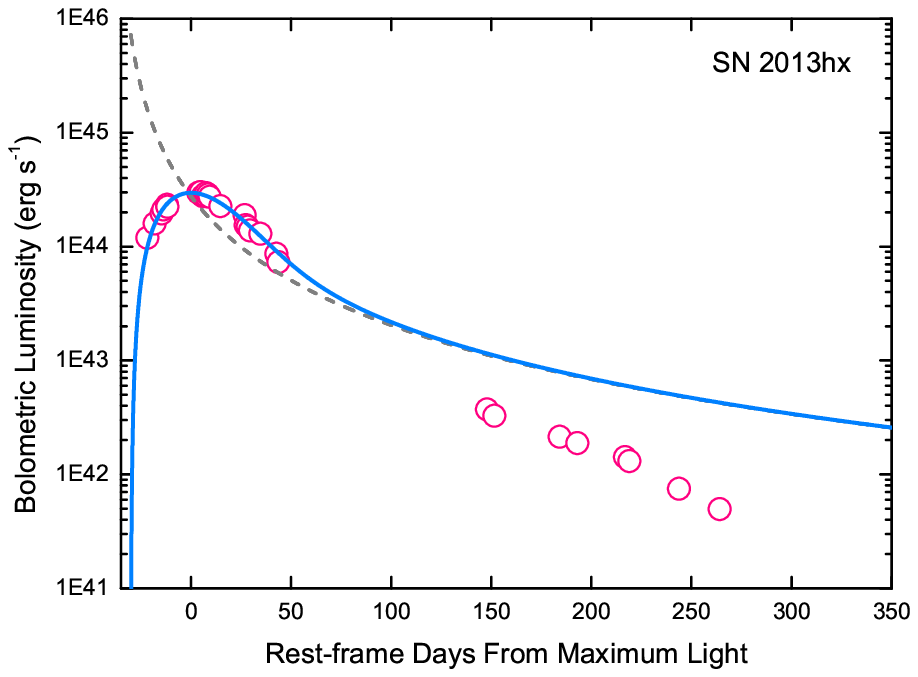}}\resizebox{0.3\hsize}{!}{\includegraphics{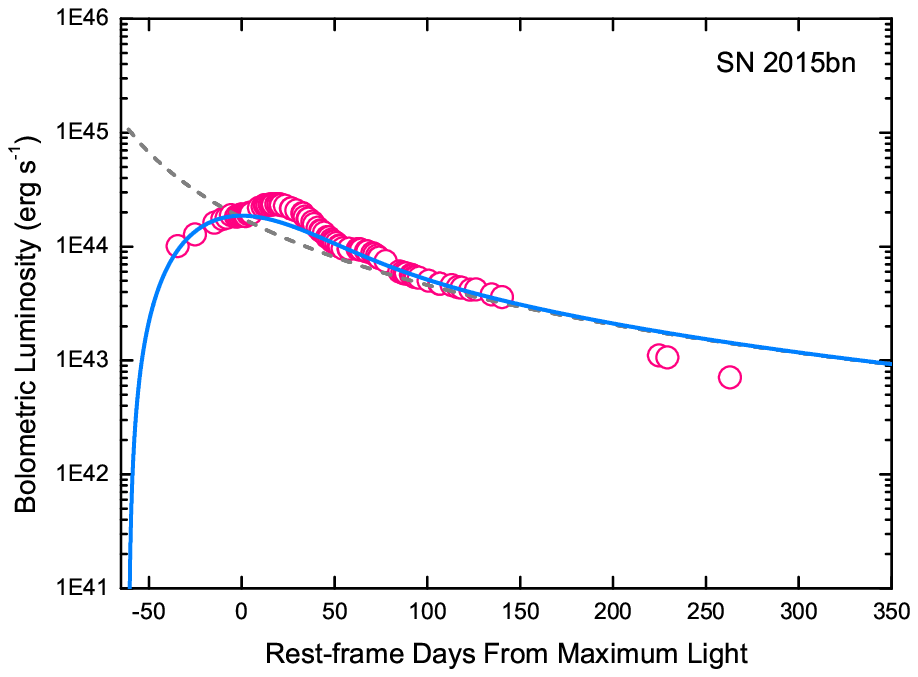}}
\center{Fig. \ref{figfittings}---Continued}
\end{figure}

\begin{figure}
\centering\resizebox{0.3\hsize}{!}{\includegraphics{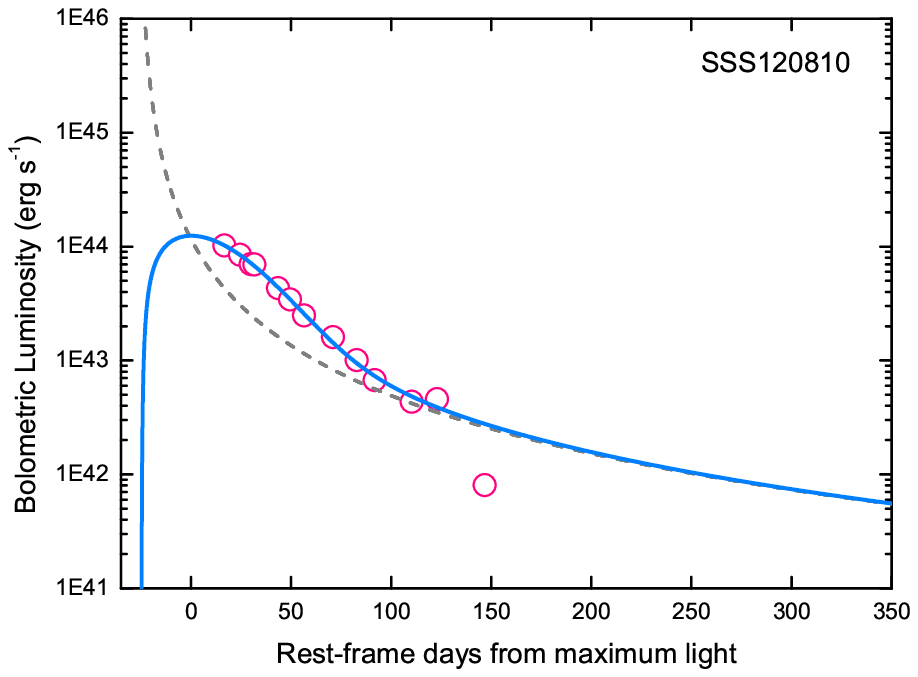}}
\center{Fig. \ref{figfittings}---Continued}
\end{figure}

\end{document}